%% file: main.tex
\documentclass[sigconf]{acmart}
\AtBeginDocument{%
  \providecommand\BibTeX{{%
    \normalfont B\kern-0.5em{\scshape i\kern-0.25em b}\kern-0.8em\TeX}}}

\copyrightyear{2023}
\acmYear{2023}
\setcopyright{acmcopyright}
\acmConference[WSDM '23] {Proceedings of the Sixteenth ACM International Conference on Web Search and Data Mining}{February 27--March 3, 2023}{Singapore, Singapore.}
\acmBooktitle{Proceedings of the Sixteenth ACM International Conference on Web Search and Data Mining (WSDM '23), February 27--March 3, 2023, Singapore, Singapore}
\acmPrice{15.00}
\acmISBN{978-1-4503-9407-9/23/02}
\acmDOI{10.1145/3539597.3570408}

\usepackage{multirow}
\usepackage{enumitem}
\usepackage{tabularx}
\newcolumntype{Y}{>{\small\centering\arraybackslash}X}
\usepackage{graphicx}
\usepackage{subcaption}

\usepackage{threeparttable}
\usepackage{algorithm}
\usepackage{algorithmic}
\usepackage[algo2e,ruled,linesnumbered]{algorithm2e}

\begin{document}

\title{DisenPOI: Disentangling Sequential and Geographical Influence for Point-of-Interest Recommendation}

\author{Yifang Qin}
\email{qinyifang@pku.edu.cn}
\orcid{0000-0002-7520-8039}
\authornote{Authors contributed equally to this research.}
\affiliation{%
  \institution{Peking University}
  \city{Beijing}
  \country{China}
}

\author{Yifan Wang}
\email{yifanwang@pku.edu.cn}
\orcid{0000-0001-7764-8698}
\authornotemark[1]
\affiliation{%
  \institution{Peking University}
  \city{Beijing}
  \country{China}}

\author{Fang Sun}
\email{fts@pku.edu.cn}
\orcid{0000-0001-5693-0379}
\authornotemark[1]
\affiliation{%
  \institution{Peking University}
  \city{Beijing}
  \country{China}}

\author{Wei Ju}
\email{juwei@pku.edu.cn}
\orcid{0000-0001-9657-951X}
\authornote{Co-corresponding authors.}
\affiliation{%
  \institution{Peking University}
  \city{Beijing}
  \country{China}}
  
\author{Xuyang Hou}
\email{houxuyang@meituan.com}
\orcid{0000-0003-0387-3547}
\affiliation{%
  \institution{Meituan}
  \city{Beijing}
  \country{China}}

\author{Zhe Wang}
\email{wangzhe65@meituan.com}
\orcid{0000-0002-0959-2714}
\affiliation{%
  \institution{Meituan}
  \city{Beijing}
  \country{China}}

\author{Jia Cheng}
\email{jia.cheng.sh@meituan.com}
\orcid{0000-0003-1702-4263}
\affiliation{%
  \institution{Meituan}
  \city{Beijing}
  \country{China}}

\author{Jun Lei}
\email{leijun@meituan.com}
\orcid{0000-0002-4015-8668}
\affiliation{%
  \institution{Meituan}
  \city{Beijing}
  \country{China}}

\author{Ming Zhang}
\email{mzhang_cs@pku.edu.cn}
\orcid{0000-0002-9809-3430}
\authornotemark[2]
\affiliation{%
  \institution{Peking University}
  \city{Beijing}
  \country{China}}








\renewcommand{\shortauthors}{Yifang Qin et al.}
\begin{abstract}
Point-of-Interest (POI) recommendation plays a vital role in various location-aware services. It has been observed that POI recommendation is driven by both sequential and geographical influences. However, since there is no annotated label of the dominant influence during recommendation, existing methods tend to entangle these two influences, which may lead to sub-optimal recommendation performance and poor interpretability.
In this paper, we address the above challenge by proposing \textbf{DisenPOI}, a novel \textbf{Disen}tangled dual-graph framework for \textbf{POI} recommendation, which jointly utilizes sequential and geographical relationships on two separate graphs and disentangles the two influences with self-supervision. The key novelty of our model compared with existing approaches is to extract disentangled representations of both sequential and geographical influences with contrastive learning. To be specific, we construct a geographical graph and a sequential graph based on the check-in sequence of a user.
We tailor their propagation schemes to become sequence-/geo-aware to better capture the corresponding influences. Preference proxies are extracted from check-in sequence as pseudo labels for the two influences, which supervise the disentanglement via a contrastive loss.
Extensive experiments on three datasets demonstrate the superiority of the proposed model.

\end{abstract}




\maketitle

\input{introduction.tex}
\input{relatedwork.tex}
\input{preliminary.tex}
\input{model.tex}

\input{experiment.tex}
\input{conclusion.tex}

\section*{Acknowledgment}
This paper is partially supported by National Key Research and Development Program of China with Grant No. 2018AAA0101902 and the National Natural Science Foundation of China (NSFC Grant Numbers 62276002 and 62106008)


\bibliographystyle{ACM-Reference-Format}
\balance
\bibliography{paper}


\end{document}

%% file: introduction.tex
\section{Introduction}

The rapid growth of Location-Based Social Networks (LBSNs) has facilitated the development of Point-of-Interest (POI) recommendation, which lies at the heart of various location-based services, including location-based advertising and online food delivery. POI recommendation is the backbone for numerous popular Apps, such as Foursquare, Gowalla and Meituan. A POI can be a restaurant, a hotel, a scenic spot, or any location-tagged item. Location is much more than an ordinary feature dimension in POI recommendation: it builds up a network structure among POIs, inter-connected by both their interactive history sequence and geographical affinity. 


Recent works on POI recommendation endeavor to integrate geographical influences with sequential transitions of user check-in data. Sequence-based methods like LSTPM ~\cite{sun2020go} combine user's long-term interest with geo-dilated short-term interest, using geographical affinity as an auxiliary in modeling sequence behavior. GeoIE ~\cite{wang2018exploiting} explicitly uses physical distance to model user-specific geographical influence and GTAG ~\cite{sun2014graph-based} performs preference propagation based on physical distance and time-slot affinity. Despite the versatility of these POI recommenders, there are rarely approaches that explicitly uncover the collaborative signals. The complex interactions among users and POIs can be readily delineated by graphs, therefore a promising approach is to leverage the rich graph structure to exploit high-order connectivity between POIs. 


Graph Neural Networks (GNNs) have been extensively deployed for recommendation to better capture the similarities between high-order neighbors. NGCF \cite{wang2019neural} and LightGCN \cite{he2020lightgcn} collects the collaborative signal on the user-POI bipartite graph, SR-GNN \cite{wu2019session} propagates along the user’s interactive sequence graph. Although these paradigms provide valuable insight as to organizing interaction data into graph structures for POI recommendation, they do not leverage geographical relationships, which is a prominent characteristic in POI recommendation.

\begin{figure}[t]
    \centering
    
    \includegraphics[width=\linewidth]{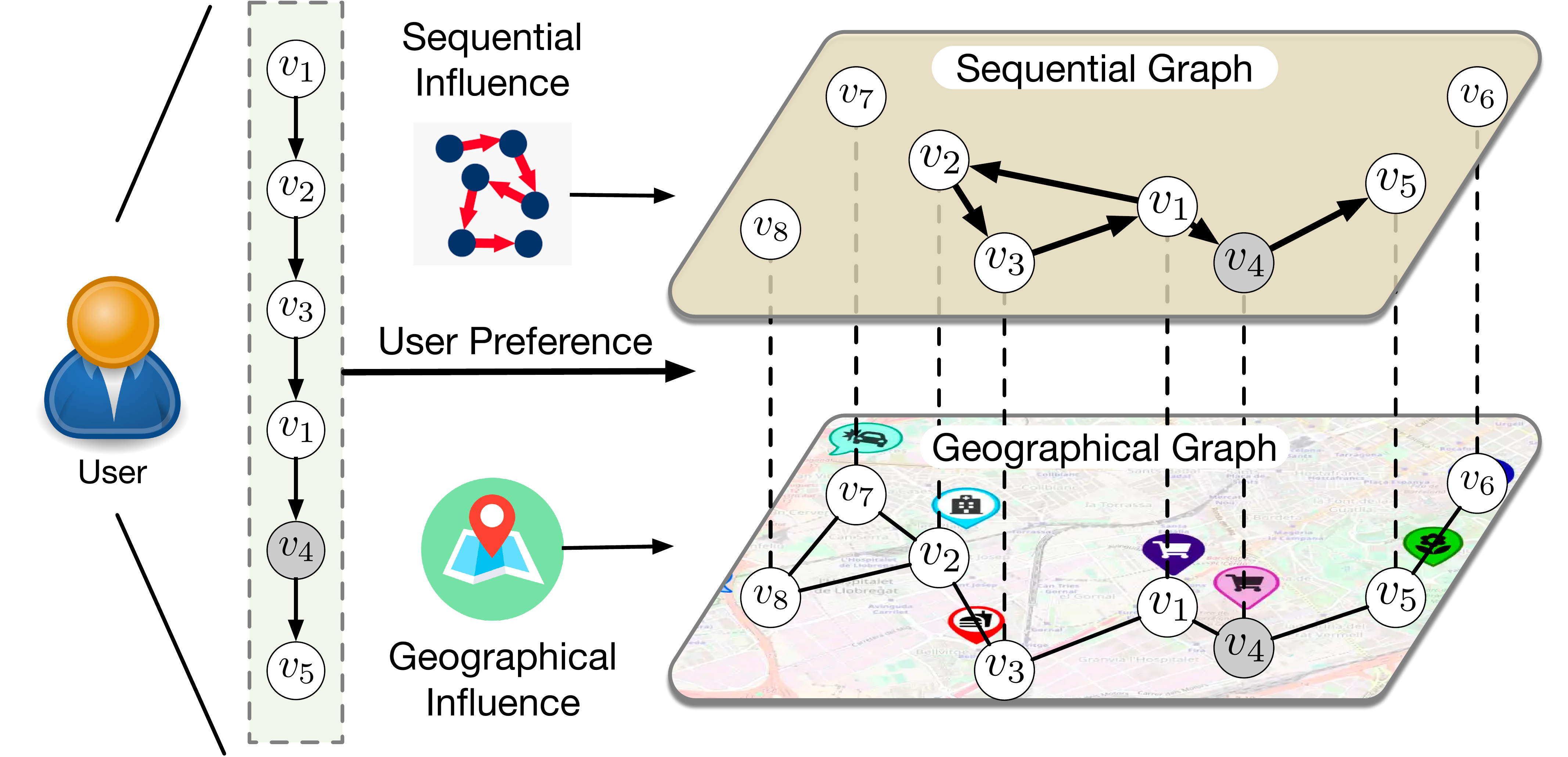}
    
    \caption{An illustration of disentangled influences behind check-ins. A visit to $v_4$ is influenced by its neighbouring nodes from geographical and sequential graph.}
    \label{fig:model_1}
\end{figure}

In fact, sequential influence and geographical influence should be regarded as equal counterparts, which are two of the major driving forces in POI recommendation. As the example in Figure \ref{fig:model_1} shows, a user's visit to $v_4$ is influenced by its neighboring nodes from geographical and sequential graphs. On the one hand, users are apt to re-visit familiar POIs that have already shown up in their interaction sequences. For example, frequent visits to clothing stores may reveal a user's preference in fashion, which indicates that a similar clothing store has more chance to be the next-to-visit POI. On the other hand, POIs in the vicinity are also more likely to be visited. For instance, a user who has just come out of a downtown shopping mall is more likely to have lunch in a nearby fast-food restaurant, rather than his/her favorite restaurant in the countryside. The two influences affect user behaviors in inherently different ways, which the previous approaches for POI recommendation do not try to distinguish. Since these approaches impose no explicit supervision on the learned user preference, there is no guarantee for the explicit capture of sequential influence and geographical influence in the user preference.
In face of the dual influences on user preference, the key is not simple combination, but rather, \textbf{disentanglement}. 
Specifically, to depict user preference at a finer granularity and enhance POI recommendation quality, we need to disentangle the two latent driving forces, thus distilling the representations for both influences provides better interpretability. 
Existing disentanglement models for recommendation either use dynamic routing (e.g. DGCF \cite{wang2020disentangled}, DisenCTR \cite{wang2022disenctr}) or self-supervised signals (e.g. MacridVAE \cite{ma2019learning}, CLSR \cite{zheng2022disentangling}) to disentangle latent interests, yet how to disentangle sequential and geographical influences remains unexplored.

In a nutshell, the ubiquity and uniqueness of the two influences in LBSNs pose two major challenges to POI recommendation: \textbf{First}, how to \textbf{depict} these two high-order connectivity influences for the user check-in data? \textbf{Second}, How to explicitly \textbf{disentangle} these two influences for POI recommendation?

To address the aforementioned challenges, we propose \textbf{DisenPOI}, a novel \textbf{Disen}tangled dual-graph framework for \textbf{POI} Recommendation, which explicitly models sequential and geographical influences on user check-in data. DisenPOI constructs two disentangled POI graphs: a geographical graph based on spatial affinity, and a sequential graph based on interaction history. To better capture high-order connectivity on different graph topologies, a distance-aware and a sequence-aware GNN are proposed to propagate on the corresponding geographical and  sequential graph. We adopt contrastive learning methods to disentangle the graph representations in a self-supervised way.
With the help of disentangled representations, DisenPOI can make recommendations that reflect both sequential and geographical influence.

We summarize our main contributions as follows:
\begin{itemize}[leftmargin=*]
\item We construct dual graphs from user's visiting interactions to jointly utilize both sequential and geographical relationships, devising sequence- and geo-aware propagation schemes for the two graphs respectively to improve embedding quality.
\item We propose to extract disentangled representations of sequential and geographical influences. Sequential and geographical proxies are obtained as self-supervised signals, which are specially tailored for the LBSN setting. To the best of our knowledge, this is the first graph-based POI recommendation framework with disentangled sequential and geographical representations.
\item We conduct extensive experiments on three real-world benchmarks for POI recommendation. Our model consistently outperforms strong baselines, achieving state-of-the-art performance on all three benchmark datasets.
\end{itemize}


%% file: relatedwork.tex
\section{Related Work}

\subsection{Location-based POI Recommendation}
Location-based POI recommendation puts special emphasis on the geographical features of POIS. GeoIE \cite{wang2018exploiting} models geographical influence between two POIs by the inner product of geo-influence and geo-susceptibility vectors. PACE \cite{yang2017bridging} combines Matrix Factorization (MF) \cite{salakhutdinov2008bayesian} approach with self-supervised learning to better exploit location information. There are also works \cite{rendle2010factorizing,cheng2013you,lian2013collaborative} that leverage Markov Chain for next-POI prediction. This line of work  \cite{liu2016predicting,feng2018deepmove,zhao2020go,sun2020go,wang2021reinforced} is characterized by their wide use of Recurrent Neural Networks (RNNs) for modeling sequential data. For example, LSTPM \cite{sun2020go} combines geographical information with sequential method by estimating the spatial-temporal similarities and finding shortcuts between POIs. There are also works \cite{cheng2012fused,gao2012gscorr,gao2015content,li2016point} that consider the social influence between users. 

\subsection{GNN-based Recommendation}
Graph Neural Networks (GNNs) like Graph Convolutional Network (GCN) \cite{kipf2016semi}, Gated Graph Neural Network (GGNN) \cite{li2016gated} and Graph Attention Network (GAT) \cite{velivckovic2017graph} have been consistently showing their potency with graph-structured data. GNNs learn the hidden representation of a graph node by passing and aggregating messages collected from its neighboring nodes. By stacking GNN layers, models\cite{luo2022dualgraph,ju2022kgnn} are empowered to exploit high-order graph connectivity. Graph structures naturally exist among users and items, which makes GNNs ideal for modeling recommendation tasks. NGCF \cite{wang2019neural} and its variant LightGCN \cite{he2020lightgcn} consider the collaborative signals on a user-item bipartite graph. DSGL \cite{chu2021dynamic} uses dynamic sequential graphs to model the dynamics of user's interaction history. Recently, works like KBGNN \cite{ju2022kernel}, GSTN \cite{wang2022graph}, GEAPR \cite{li2021you} and IMUP \cite{wang2020incremental} propose to integrate geographical factors into GNN-based recommendation to improve performance under location-based settings. However, every observed interaction can be driven by various ambient influences, as well as diverse user interests. Representations learned by GNNs cannot readily uncover these rich semantics.

\subsection{Disentangled Learning in Recommendation}
Disentangled Representation Learning \cite{bengio2013representation} aims to learn separate explanatory representations from different latent factors behind data. MacridVAE \cite{ma2019learning} employs Variational Auto-Encoders (VAEs) to encode items via disentangled prototypes. CLSR \cite{zheng2022disentangling} attempts to disentangle users' long and short-term interests from their clicking sequences. GNNs are also used in disentangled recommenders, which propose to learn disentangled representation from various graph structures. Such models include DGCF \cite{wang2020disentangled} (user-item graph),  DisenHAN \cite{wang2020disenhan} (heterogeneous graph) and KGIN \cite{wang2021learning} (knowledge graph). Yet, the aforementioned works only perform disentangled learning on one type of graph topology. They are not intended to capture influences from POI locations, either. These short-comings shall be addressed by our proposed DisenPOI model, which uses two disentangled graphs to utilize sequential and geographical relationships, and extracts the rich semantics behind observed interactions by disentangling sequential and geographical influences.

%% file: preliminary.tex
\section{Preliminary}
\subsection{Problem Formulation}
Location-based Click Through Rate (CTR) prediction aims to leverage geographical information in user's interactive history with POIs. We denote the set of users as  $\mathcal{U}=\{u_1,u_2,...,u_{|\mathcal{U}|}\}$, and the set of POIs as $\mathcal{V}=\{v_1,v_2,...,v_{|\mathcal{V}|}\}$, where each POI $v\in\mathcal{V}$ is geo-encoded by its coordinates $(longitude, latitude)$ tuple, i.e. $(lon_v, lat_v)$. For each user $u\in\mathcal{U}$, his/her check-in sequence is organized into a list $s_u=[v_{u,1},v_{u,2},...,v_{u,t-1}]$, containing the $t-1$ POIs he/she has checked in, sorted by timestamp. Given user $u$ and target POI $v$, the goal of location-based CTR prediction is to predict the probability for $u$ to visit $v$, formulated as $\hat{y}_{uv}=f(u,v|s_u;\theta)$, where $f$ is a learnable function parameterized by $\theta$.

\subsection{Construction of Two Disentangled Graphs}
As we aim to make predictions based on both the geographical relationship between POIs and the sequential dependency in users' check-in history, two POI graphs with different propagation schemes are thus constructed to meet the two ends respectively.

\subsubsection{Geographical Graph}
\label{sec:geograph}
The undirected geographical graph $\mathcal{G}_g=\{\mathcal{V}, \mathcal{E}_g, A_g\}$ is built upon the location of POIs. An undirected edge $e_g=(v_i,v_j)\in\mathcal{E}_g$ indicates that the distance between $v_i$ and $v_j$ is within a specific distance threshold $\Delta d$. The edge weight matrix $A_g(i,j)$ is the geographical distance between POI $v_i$ and $v_j$. The construction of $\mathcal{G}_g$ is based on the fact that a user would usually prefer a nearby POI for the next visit.

\subsubsection{Sequential Graph}
Inspired by previous work on session-based recommendation \cite{wu2019session}, we propose to construct sequential graphs for visiting sequences. Given a user $u$ and his/her check-in history $s_u$, a directed sequential graph $\mathcal{G}_{s,u}=\{\mathcal{V}_{s,u},\mathcal{E}_{s,u}\}$ can then be constructed. For each edge $e_s=\left\langle v_i,v_j\right\rangle\in\mathcal{G}_s$, it represents that the user has successively check-in to $v_j$ after $v_i$ in $s_u$. Thus, $\mathcal{G}_{s,u}$ gleans sequential information from the check-in history.

%% file: model.tex
\section{The Proposed Model}

\subsection{Overview}
\begin{figure*}
    \centering
    \includegraphics[width=0.9\linewidth]{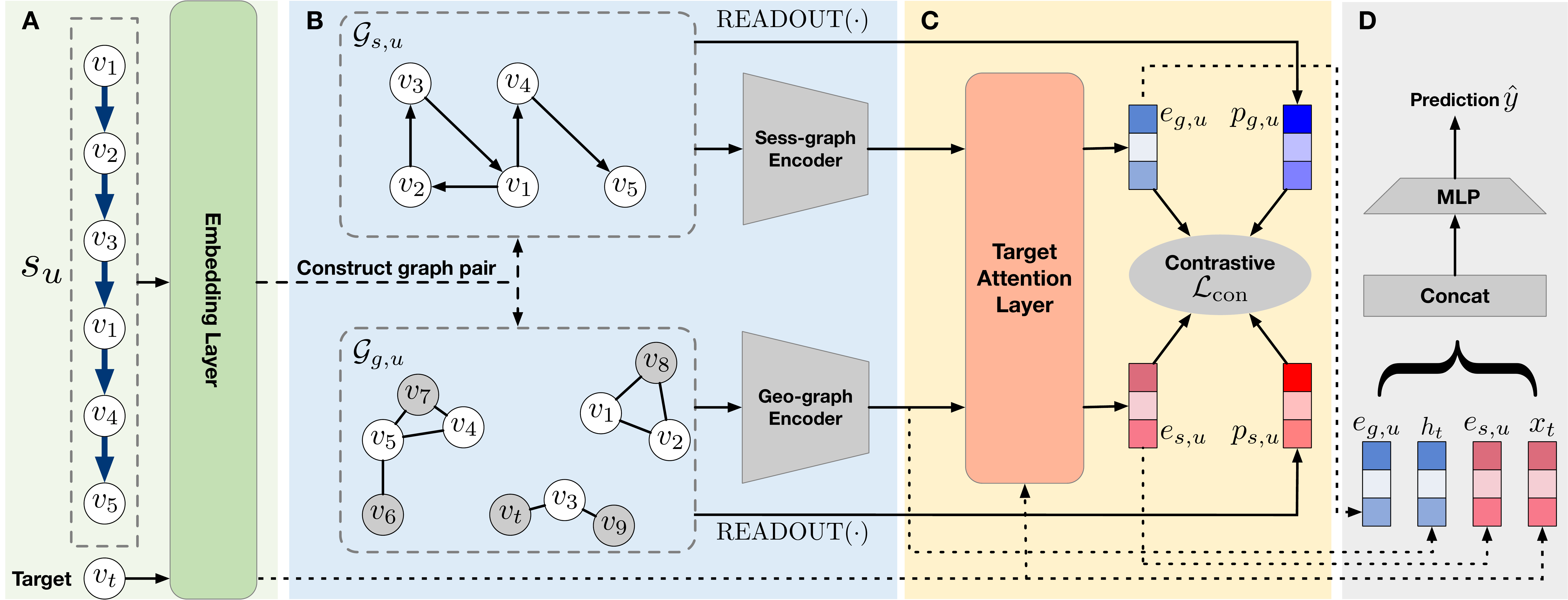}
    \caption{Illustration of DisenPOI. A) Embedding Layer for POIs; B) graph propagation layer that generates disentangled representations $e_{g,u},e_{s,u}$ and their proxies $p_{g,u},p_{s,u}$; C) soft-attention layer that aggregates POI embeddings for contrastive loss and model prediction; D) prediction layer to make predictions.}
    \label{fig:model_2}
\end{figure*}

The proposed model has four modules as illustrated in Figure \ref{fig:model_2}. Given a user's check-in history and a target POI, the model first maps IDs of the involved POIs into a hidden space, i.e. $X\in\mathbb{R}^{D\times D}$, to get their initial embeddings. After the POIs are mapped, the model constructs two graphs, $G_g$ and $G_s$, which reflect the geographical and sequential features of the given visiting history respectively. Two graph propagation modules are used upon the drawn graphs to generate representations from a disentangled view. A soft-attention mechanism is applied to extract the user preference behind visits on both modules, in order to dynamically measure the influence of visited and nearby POIs on the target POI. To explicitly disentangle this hybrid influence into two independent fields, 
we introduce a contrastive auxiliary loss to ensure that each module is able to extract disentangled information from the corresponding POI graph. Finally, we use an MLP layer to aggregate these disentangled representations and generate CTR prediction for the target POI.

During the training process, we adopt a curriculum learning scheme to integrate the contrastive loss with the target function. We control the curriculum weight to train the model in an easy-to-difficult manner so that our model can be trained adequately to give valid predictions while maintaining effective disentanglement.

\subsection{Propagation on Disentangled Graphs}
The graph propagation layer encodes the constructed graphs and outputs two sets of POI representations with geographical and sequential information. 
Since the two disentangled graphs are organized according to sequential/geographical relationships, we apply sequence-/geo-aware graph propagation schemes on them respectively, to better exploit the intrinsic characteristics of each graph. 

\subsubsection{Geographical Graph Propagation Layer}

Given the location-based graph $\mathcal{G}_g$, we build our geographical graph propagation layer using the message-passing scheme of GNNs~\cite{hamilton2017inductive,xu2018representation}.

\textbf{Message Construction.}
For a pair of neighboring POIs $v_i, v_j$, the message on the $l$-th layer is defined as:
\begin{equation}
\label{eqt:1}
    m^{(l)}_{j\leftarrow i}=f_d(h_i^{(l-1)},h^{(l-1)}_j),
\end{equation}
where $f_d(\cdot)$ is the message encoding function that takes item representation $h^{(l-1)}$ from the previous GNN layer. Specifically, the input of the first layer is initialized with POI's embedding $X$: \begin{equation}
    h_i^{(0)}=x_i, \forall v_i\in\mathcal{V}.
\end{equation}
In order to better leverage the geographical information in $\mathcal{G}_g$, the message function $f_d(\cdot)$ is expected to reflect the distance influence between POIs. Specifically, POIs that are closer tend to share more similarities than distant ones. Formally, $f_d(\cdot)$ is implemented as:
\begin{equation}
\label{eqt:3}
    m_{j\leftarrow i}^{(l)}=\frac{1}{\sqrt{|\mathcal{N}_i||\mathcal{N}_j|}}(W_1^{(l)}h_i^{(l-1)}+w(d_{ij})W_2^{(l)}h_i^{(l-1)}\odot h_j^{(l-1)}),
\end{equation}
where $W_1,W_2\in\mathbb{R}^{D\times D}$ are trainable weight matrices as linear transformations for lower layer representations. To take the effect of distance into consideration, an element-wise production item is included. Distance kernel $w(d_{ij})=e^{-d_{ij}^2}$ decays exponentially as $d_{ij}$, the distance between $v_i$ and $v_j$ grows. We use the graph Laplacian norm like GCNs~\cite{kipf2016semi}, where $\mathcal{N}_i$ and $\mathcal{N}_j$ denote the first-hop neighbors for $v_i$ and $v_j$ on $\mathcal{G}_g$ respectively.

\textbf{Message Aggregation.}
To generate node representations on each GNN layer, the message aggregation function is defined as:
\begin{equation}
    h_j^{(l)}=\text{LeakyReLU}(m_{j\leftarrow j}+\sum_{i\in\mathcal{N}_u}m_{j\leftarrow i}),
\end{equation}
where $m_{j\leftarrow j}=W_1h_j^{(l-1)}$ is induced by the representation of $v_j$. 

After propagating along $L$ GNN layers, we adopt the hidden representation of the $L$-th layer as the geographical encoding of all the POIs in $\mathcal{V}$:
\begin{equation}
H_g=[h_1^{(L)};h_2^{(L)};...;h_{|\mathcal{V}|}^{(L)}].
\end{equation}

As for user $u$ with check-in history $s_u$, we assume that his/her geographical preference can be captured by aggregating the high-order geographical neighbors of POIs in his/her check-in history. In practice, the user's geographical encoding $H_{g,u}$ is the list of the geographical embeddings of all the POIs that appear in $s_u$.  

\subsubsection{Sequential Graph Propagation Layer}
Given a target user $u$ and the constructed sequential graph $\mathcal{G}_{s,u}$, we obtain the hidden representation of each item in the user's visiting history to exploit sequential influence. A user's visit sequence contains not only information about his preferences on POIs, but also the evolving history of his interests, which implies his tendency for the next POI to visit. Since plain GCNs are sequential-invariant, we consider adopting a graph propagation method that can fully leverage sequential information in the session graph. For the given reasons, we apply Gated Graph Neural Networks (GGNNs)~\cite{li2016gated} on $\mathcal{G}_{s,u}$ to better leverage the sequential information. For each node $v_i\in\mathcal{G}_{s,u}$ and its embedding $x_i$, the hidden representation of $v$ is updated as:

\begin{align}
\begin{split}
    h_v^{(1)}&=[x_i^\top,0]^\top
\end{split}\\
\begin{split}
    a_{v}^t&=A_{v:}^\top[h_1^{(t-1)\top},...,h_{|\mathcal{V}|}^{(t-1)\top}]^\top+b
\end{split}\\
\begin{split}
    z_v^t&=\sigma(W^za_v^{(t)}+U^zh_v^{(t-1)})
\end{split}\\
\begin{split}
    r_v^t&=\sigma(W^ra_v^{(t)}+U^rh_v^{(t-1)})
\end{split}\\
\begin{split}
    \widetilde{h_v^{(t)}}&=\tanh(W_oa_v^{(t)}+U_o(r_v^t\odot h_v^{(t-1)}))
\end{split}\\
\begin{split}
    h_v^{(t)}&=(1-z_v^t)\odot h_v^{(t-1)}+z_v^t\odot\widetilde{h_v^{(t)}},
\end{split}
\end{align}
where connection matrix $A_v\in\mathbb{R}^{|\mathcal{V}|\times2|\mathcal{V}|}$ determines the connectivity of nodes in $\mathcal{G}_{s,u}$, $W$s and $U$s are trainable parameters. Inspired by Gate Recurrent Unit (GRU), GGNN proceeds on all nodes of the sequential graph sequentially. The propagation of hidden states is controlled by an update gate $z^t_v$ and a reset gate $r^t_v$. The gated controlled propagation on $\mathcal{G}_{s,u}$ models the sequential effect behind user behavior. We obtain the output of the last hidden layer as the sequential encoding of POIs:
\begin{equation}
\label{eqt:endprop}
    H_{s,u}=[h_{u,1}^{(t)},h_{u,2}^{(t)},...,h_{u,|s_u|}^{(t)}]
\end{equation}

\subsection{Soft-attention Mechanism}

After encoding the two constructed graphs from user's previous visiting history, we introduce a soft-attention mechanism to better aggregate the encodings according to current target POI $v_t$. For geographical graph encoding $H_{g,u}=[h_1,h_2,...,h_{|s_u|}]$, we denote $v_t$'s hidden representation $h_t$ as the query, the soft-attention mechanism to generate geographical embedding $e_{g,u}$ is formulated as:
\begin{align}
    w_{i}&=\alpha_g^\top\sigma(Q_gh_{t}+K_gh_i) \\
    e_{g,u}&=\sum_{i=1}^{|s_u|}w_ih_i,
\label{eqt:geo}
\end{align}
where $\sigma$ denotes sigmoid function, the attention vector $\alpha_g\in\mathbb{R}^D$, query and key matrix $Q_g,K_g\in\mathbb{R}^{D\times D}$ are trainable parameters. Similarly, for $H_{s,u}=[h_1',h_2',...,h_{|s_u|}']$, we take $v_t$'s initial embedding $x_t$ as query vector to generate sequential embedding $e_{s,u}$:
\begin{align}
    w_{i}'&=\alpha_s^\top\sigma(Q_sx_t+K_sh_i') \\
    e_{s,u}&=\sum_{i=1}^{|s_u|}w_i'h_i',
\label{eqt:sess}
\end{align}

\subsection{Self-supervised Disentanglement}

Since the sequential and geographical effects have separate influences on next-to-visit POI, it is essential to disentangle the two representations, $e_{g,u}$ and $e_{s,u}$, from each other so that model could fully utilize the information for CTR prediction. Inspired by previous work on self-supervised disentanglement learning~\cite{zheng2022disentangling}, we suggest obtaining proxies for both embeddings. Specifically, two readout functions are applied to the outputs of the graph propagation layer. We choose mean pooling as readout function to apply to both geographical neighbors and user's previous visited POIs to generate two proxies for $e_{g,u}$ and $e_{s,u}$. Formally, we have:
\begin{align}
    p_{g,u}&=\text{READOUT}(\{x_j|v_j\in\mathcal{N}_{s_u}\})=\frac{1}{\sum_{v_i\in s_u}|\mathcal{N}_i|}\sum_{v_i\in s_u}\sum_{v_j\in\mathcal{N}_i}x_j 
\label{eqt:geo_prox}\\
    p_{s,u}&=\text{READOUT}(\{x_i|v_i\in s_u\})=\frac{1}{|s_u|}\sum_{v_i\in s_u}x_i,
\label{eqt:sess_prox}
\end{align}
which conduct mean pooling on $s_u$'s one-hop geographical neighbors 
$\mathcal{N}_{s_u}$ and $s_u$ respectively. Since $p_{g,u}$ and $p_{s,u}$ are readout from two different graphs, it's reasonable to assume that each proxy carries representative information of its corresponding graph features. To be specific, we assume that $p_{g,u}$ carries abundant information about the user's geographical preference, while $p_{s,u}$ carries information about the user's historical interests.

Before the embeddings and their proxies were sent to generate self-supervising signals, we design projection heads for each disentangled representations following previous work~\cite{chen2020simple}. To formulate, the projected representations are calculated as:
\begin{align}
    e_{g,u}'=\text{proj}_g(e_{g,u}),\ p_{g,u}'=\text{proj}_g(p_{g,u})\\
    e_{s,u}'=\text{proj}_s(e_{s,u}),\ p_{s,u}'=\text{proj}_s(p_{s,u}),
\end{align}
where $\text{proj}_g,\text{proj}_s:\ \mathbb{R}^D\rightarrow\mathbb{R}^D$ are linear transformations with trainable parameters. The projection heads project the embeddings into another latent space where information from different aspects can be easily clarified.

We treat the projected pooling representations, $p_{g,u}'$ and $p_{s,u'}$ as hidden proxies for $e_{g,u}'$ and $e_{s,u}'$, which brings supervising signals for self-supervised disentanglement. In other words, the two proxies serve as positive samples for the corresponding embedding, while serving as negative samples for each other. With the above assumption, we introduce the following contrastive loss:
\begin{equation}
    \mathcal{L}_{\text{con}}=f(e_{g,u}', p_{g,u}', p_{s,u}')+f(e_{s,u}', p_{s,u}', p_{g,u}'),
\label{eqt:con}
\end{equation}
where $f(\cdot)$ denotes the Bayesian
Personalized Ranking loss~\cite{rendle2012bpr}. Formally, we have:
\begin{equation}
    f(a,p,q)=\text{Softplus}(\left\langle a,q\right\rangle-\left\langle a,p\right\rangle),
\end{equation}
where $\left\langle\cdot,\cdot\right\rangle$ denotes the inner product of two given representations. 
\subsection{Location-based CTR Prediction Layer}

\subsubsection{Model Prediction}
After obtaining the disentangled embeddings of the user's visiting history $e_{g,u}$ and $e_{s,u}$, we concatenate the embeddings with target $v_t$'s embedding $x_t$ and geographical representation $h_{t}$. A 2-layer MLP is applied to predict the click-through rate of $v_i$, i.e. $\hat{y}$, which is a common practice in CTR prediction~\cite{zhou2018deep,zhou2019deep}. Formally, $\hat{y}$ is calculated by:
\begin{equation}
\label{eqt:pred}
    \hat{y}=\sigma(\text{MLP}(\text{CONCAT}(e_{g,u},e_{s,u},x_{t},h_{t})))
\end{equation}

\subsubsection{Target Function and Curriculum Learning}
Given the label $y$, we adopt binary cross-entropy loss for supervised CTR prediction, which is formulated as:
\begin{equation}
    \mathcal{L}_{\text{rec}}=-\sum_{(u,v)}y\log\hat{y}+(1-y)\log(1-\hat{y}).
\label{eqt:rec}
\end{equation}
After adding the disentangled constraint to the model, we have the overall model loss:
\begin{equation}
    \mathcal{L}=\mathcal{L}_{\text{rec}}+\beta * \mathcal{L}_{\text{con}},
\label{eqt:tot_loss}
\end{equation}
where $\beta$ is the weight of disentanglement. To optimize the model, we propose a curriculum learning method to have the training process follow an easy-to-difficult process. To be specific, we train the model with a warm-up procedure by dynamically increasing the weight of the contrastive loss:
\begin{equation}
    \beta=\max\{\alpha, \gamma * k\},
\end{equation}
where $\alpha$ and $\gamma$ are hyper-parameters, $k$ denotes the current curriculum number. Under curriculum training, the strength of disentanglement would increase to the maximum. 

%% file: experiment.tex
\section{Experiment}
In this section, we conduct comprehensive experiments on three real-world location-based recommendation datasets to answer the following research questions:
 

\textbf{RQ1:} How does the proposed DisenPOI perform compared with the current state-of-art baseline methods on location-based CTR predictions? How can DisenPOI alleviate the data sparsity issue with the help of disentangled information?

\textbf{RQ2:} Is the idea of disentangling geographical and sequential influence necessary to achieve a promising model performance? What are the influences of the different hyper-parameters?


\textbf{RQ3:} How effectively does DisenPOI disentangle POI representations? Can DisenPOI recommend POIs with marked geographical structure as expected?

\subsection{Experimental Setup}
\begin{table}
\centering
\caption{Descriptive statistics of our three datasets.}
\label{tab:1}
\label{tab:statics}
\setlength{\tabcolsep}{3pt}
\begin{tabular}{c|cccc} 
\toprule 
\textbf{Dataset} & \#User & \#POI & Interactions & Avg.SeqLen \\
\midrule 
Tokyo & 2,293 & 61,858 & 573,703 & 250.20 \\
\midrule
New York & 1,083 & 38,333 & 227,428 & 210.00 \\
\midrule
Meituan & 164,855 & 5,077 & 470,095 & 2.85 \\
\bottomrule 
\end{tabular}
\end{table}

\subsubsection{Evaluation Datasets}
We conduct experiments on three real-world datasets collected from two platforms, namely \textbf{Foursquare}~\cite{yang2014modeling} and \textbf{Meituan} to evaluate the proposed model. The detailed statistics of the three datasets are presented in table~\ref{tab:1}.

\begin{enumerate}[leftmargin=*]
    \item \textbf{Foursquare}\footnote{{\url{https://sites.google.com/site/yangdingqi/home/foursquare-dataset}}}: It contains two subsets, which are users' check-in records collected in Tokyo and New York City respectively from 12 April 2012 to 16 February 2013. There are 573,703 check-ins in Tokyo dataset and 227,428 check-ins in New York City (NYC).
    \item \textbf{Meituan}: It is a dataset drawn from Meituan users' check-ins to restaurants from 7 July 2021 to 20 July 2021. We collect the records in Shanghai, which has 470,095 check-ins, with 164,855 users and 5,077 POIs included.
\end{enumerate}

We sort the recorded user interactions in each dataset in chronological order. All visited POIs and the corresponding previous visiting subsequences make up the positive samples. For each positive sample, we randomly select one of the user's unvisited POIs as the negative sample. The last interaction of each user is reserved for evaluation, while the remaining part is used for training. The evaluation set is then randomly split into two equal-sized subsets as test and validation set respectively.

\subsubsection{Baselines}
To demonstrate the effectiveness of the proposed DisenPOI, we compare it with three classes of baselines: (A) Traditional sequence-based methods; (B) Graph-based methods; and (C) Location-based methods which leverage geographical information. 
\begin{itemize}[leftmargin=*]
    \item \textbf{DIN}\cite{zhou2018deep} (A): a classical method for sequence-based CTR prediction that uses target attention for user's context embedding.
    \item \textbf{DIEN}\cite{zhou2019deep} (A): a variant of DIN which leverages GRUs to capture the evolution of user interest.
    \item \textbf{SR-GNN}\cite{wu2019session} (B): a graph-based method that models user's interaction history as session graphs.
    \item \textbf{NGCF}\cite{wang2019neural} (B): a graph-based method that applies GNN on the user-item interaction graph.
    \item \textbf{LightGCN}\cite{he2020lightgcn} (B): a variant of NGCF. It's one of the state-of-art methods for graph-based recommendation.
    \item \textbf{GeoIE}\cite{wang2018exploiting} (C): one of the state-of-art location-based recommendation methods, which takes the influence of geographical location and distance into consideration.
    \item \textbf{LSTPM}\cite{sun2020go} (C): an LSTM-based method that proposes the geo-dilated LSTM to model geographical effect.
    \item \textbf{GSTN}\cite{wang2022graph} (C): a location-based graph recommendation model that optimizes the spatial-temporal similarities between POIs.
\end{itemize}

Since we have constructed training and evaluation set with positive and negative samples, we adopt AUC and Logloss as evaluation metrics, which is a common practice in CTR prediction \cite{zhou2018deep,zhou2019deep}. The experiment results on three datasets are shown in table \ref{tab:2}.

\subsubsection{Implementation Detail}
We implement our DisenPOI and all baselines in Pytorch. The embedding size is fixed to 64. While some of the above baseline models are not intended for CTR prediction, we apply the same-sized 2-layer MLP on the backend of each model to compare their performances on CTR prediction on an equal basis. For DisenPOI, we set hyper-parameters $\alpha=0.2, \gamma=0.004$ for training and evaluation and we use 2 layers of GNNs on both graph modules. The distance threshold $\Delta d$ is set to 1km when constructing geographical graphs. All models are optimized with Adam optimizer with the learning rate $lr=0.001$. Our implementation is publicly available at https://github.com/Yifang-Qin/DisenPOI

\subsection{Performance Comparison (RQ1)}

\subsubsection{Overall Comparison}
\begin{table}
\centering
\caption{The performance of DisenPOI (Ours) and other baseline methods over three datasets. $\star$ indicates DisenPOI outperforms the best POI at p-value<0.05 level, unpaired t-test.}
\label{tab:2}
\begin{tabularx}{\linewidth}{cYYYYYY}
\toprule 
\multirow{2}{*}{Model} & \multicolumn{2}{c}{Tokyo} & \multicolumn{2}{c}{NYC} & \multicolumn{2}{c}{Meituan}\\ 
\cmidrule[0.5pt](lr){2-3}\cmidrule[0.5pt](lr){4-5}\cmidrule[0.5pt](lr){6-7}
& AUC$\uparrow$ & Logloss$\downarrow$ & AUC & Logloss & AUC & Logloss\\
\midrule 
DIN & 0.8623 & \underline{0.4174} & 0.8015 & 0.5203 & 0.5596 & 0.6920 \\
DIEN & 0.8643 & 0.4544 & 0.8062 & 0.5582 & 0.5591 & 0.6891 \\
\midrule
SR-GNN & 0.8867 & 0.4481 & 0.8491 & 0.5262 & 0.5751 & \underline{0.6844} \\
NGCF & 0.8899 & 0.4436 & 0.8461 & \underline{0.4788} & 0.5698 & 0.6857 \\
LightGCN & 0.8998 & 0.4348 & 0.8729 & 0.4815 & 0.5682 & 0.6859 \\
\midrule
GeoIE & \underline{0.9083} & 0.4203 & \underline{0.8769} & 0.4810 & 0.5431 & 0.6926 \\
LSTPM & 0.8745 & 0.4374 & 0.8568 & 0.4821 & 0.5637 & 0.6867 \\
GSTN & 0.8906 & 0.4312 & 0.8521 & 0.5323 & \underline{0.5807} & 0.6845 \\
\midrule
Ours & \textbf{0.9246\textsuperscript{$\star$}} & \textbf{0.3608\textsuperscript{$\star$}} & \textbf{0.8941\textsuperscript{$\star$}} & \textbf{0.4248\textsuperscript{$\star$}} & \textbf{0.5823\textsuperscript{$\star$}} & \textbf{0.6836\textsuperscript{$\star$}} \\
\bottomrule 
\end{tabularx}
\end{table}

As shown in table \ref{tab:2}, methods that leverage geographical information (GeoIE, LSTPM, GSTN, and DisenPOI) generally outperform other models, which is a reasonable result since the high relativity between user's check-in history and POIs' geographical locations in location-based recommendation tasks. 
The proposed DisenPOI outperforms all of the state-of-art baselines with a significant improvement. In particular, the testing AUC is improved over the strongest baseline w.r.t AUC by 1.7\%, 2.0\%, 0.3\%; Logloss by 1.3\%, 1.1\%, 0.2\% in \textbf{Foursquare Tokyo}, \textbf{Foursquare NYC} and \textbf{Meituan} respectively. The result shows that DisenPOI achieves the best performance on CTR prediction task.

\subsubsection{Performance on Cold-start Recommendation}
\begin{figure}
\centering
\setlength{\abovecaptionskip}{0.1cm}
\setlength{\belowcaptionskip}{-0.3cm}
\begin{subfigure}{0.45\linewidth}
    \setlength{\abovecaptionskip}{-0.0cm}
    \setlength{\belowcaptionskip}{-0.0cm}
    \includegraphics[width=\linewidth]{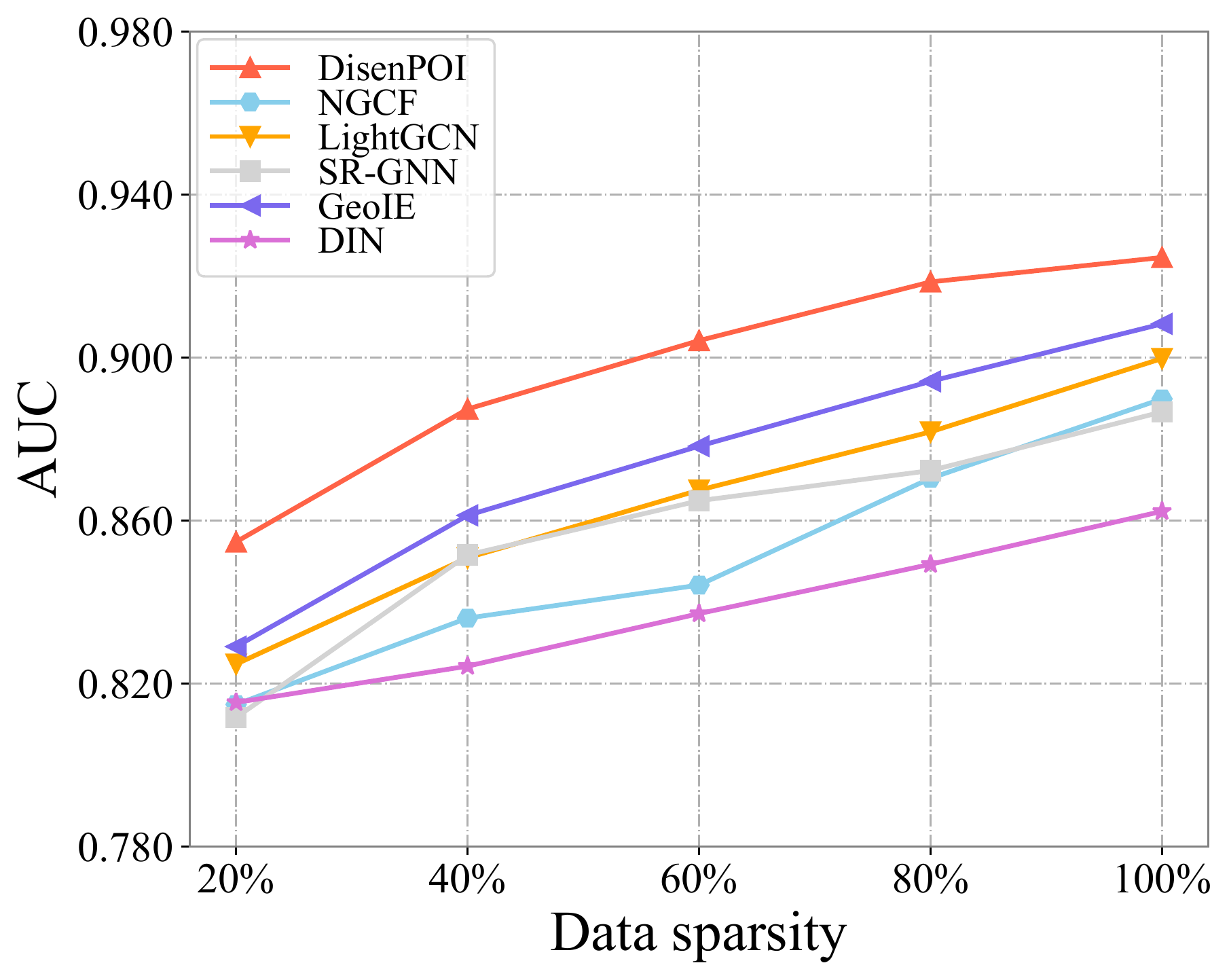}
    \caption{AUC on \textbf{Tokyo}}
    \label{fig:colda}
\end{subfigure}
\begin{subfigure}{0.45\linewidth}
    \setlength{\abovecaptionskip}{-0.0cm}
    \setlength{\belowcaptionskip}{-0.0cm}
    \includegraphics[width=\linewidth]{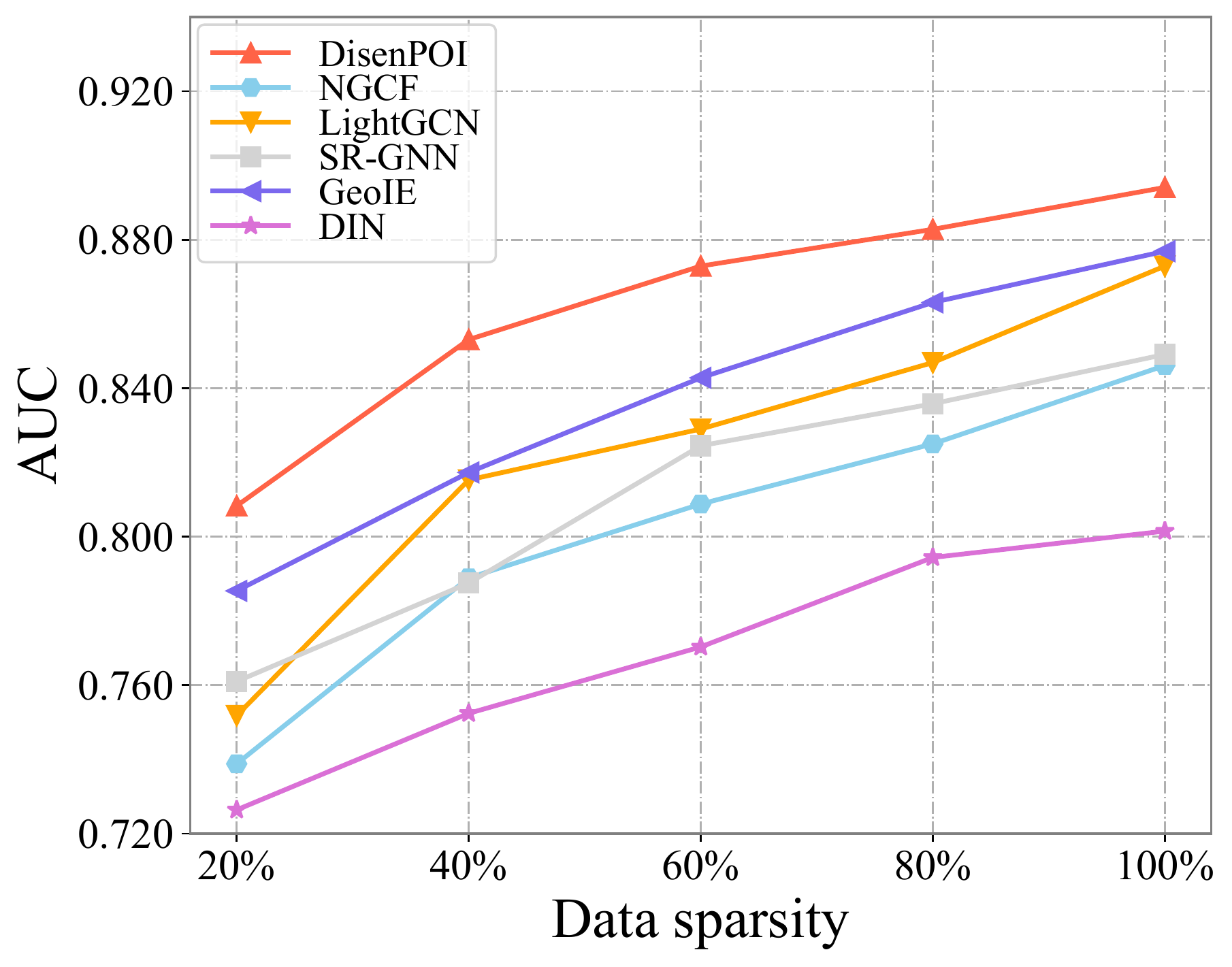}
    \caption{AUC on \textbf{NYC}}
    \label{fig:coldb}
\end{subfigure}

\begin{subfigure}{0.45\linewidth}
    \setlength{\abovecaptionskip}{-0.0cm}
    \setlength{\belowcaptionskip}{-0.0cm}
    \includegraphics[width=\linewidth]{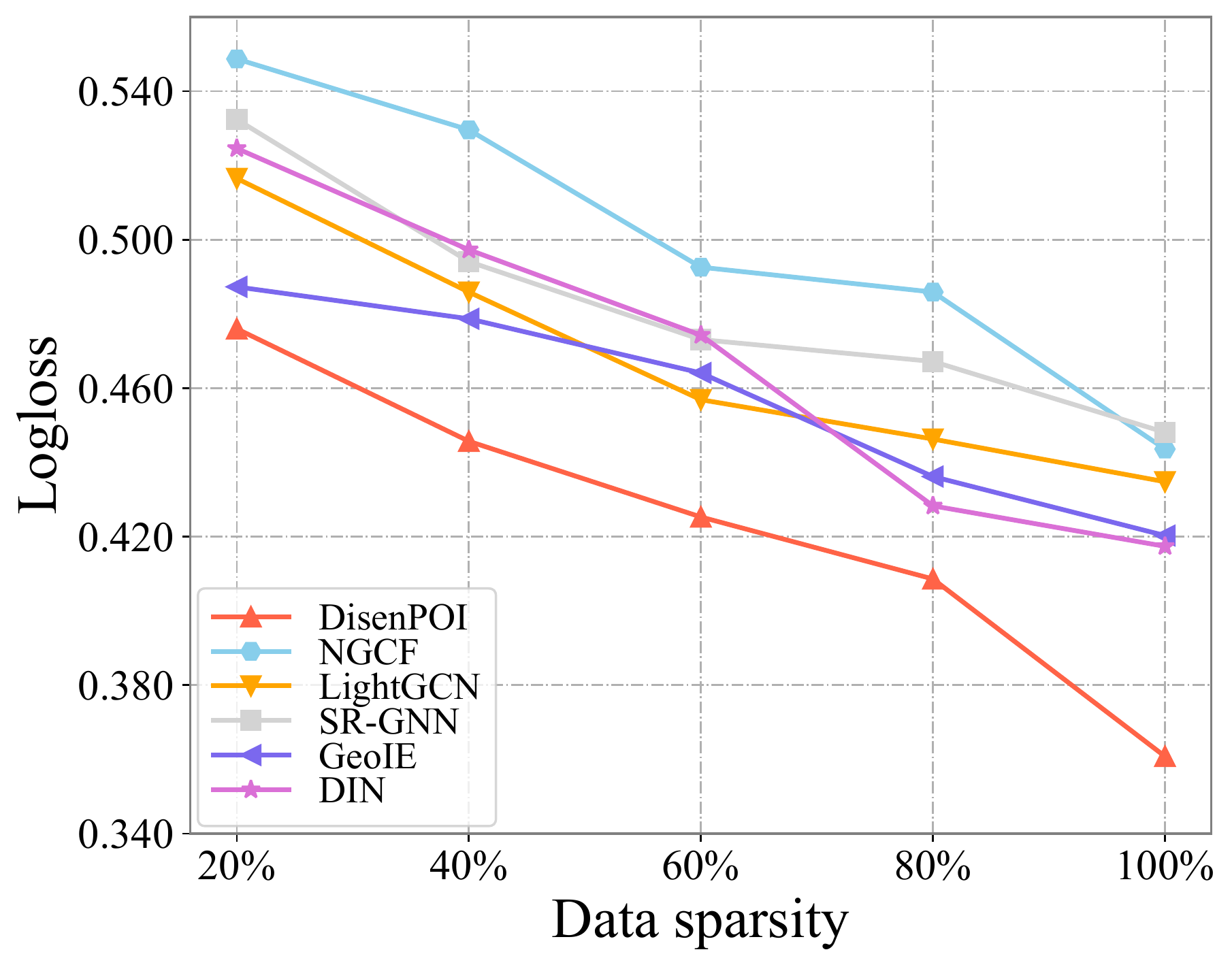}
    \caption{Logloss on \textbf{Tokyo}}
    \label{fig:coldc}
\end{subfigure}
\begin{subfigure}{0.45\linewidth}
    \setlength{\abovecaptionskip}{-0.0cm}
    \setlength{\belowcaptionskip}{-0.0cm}
    \includegraphics[width=\linewidth]{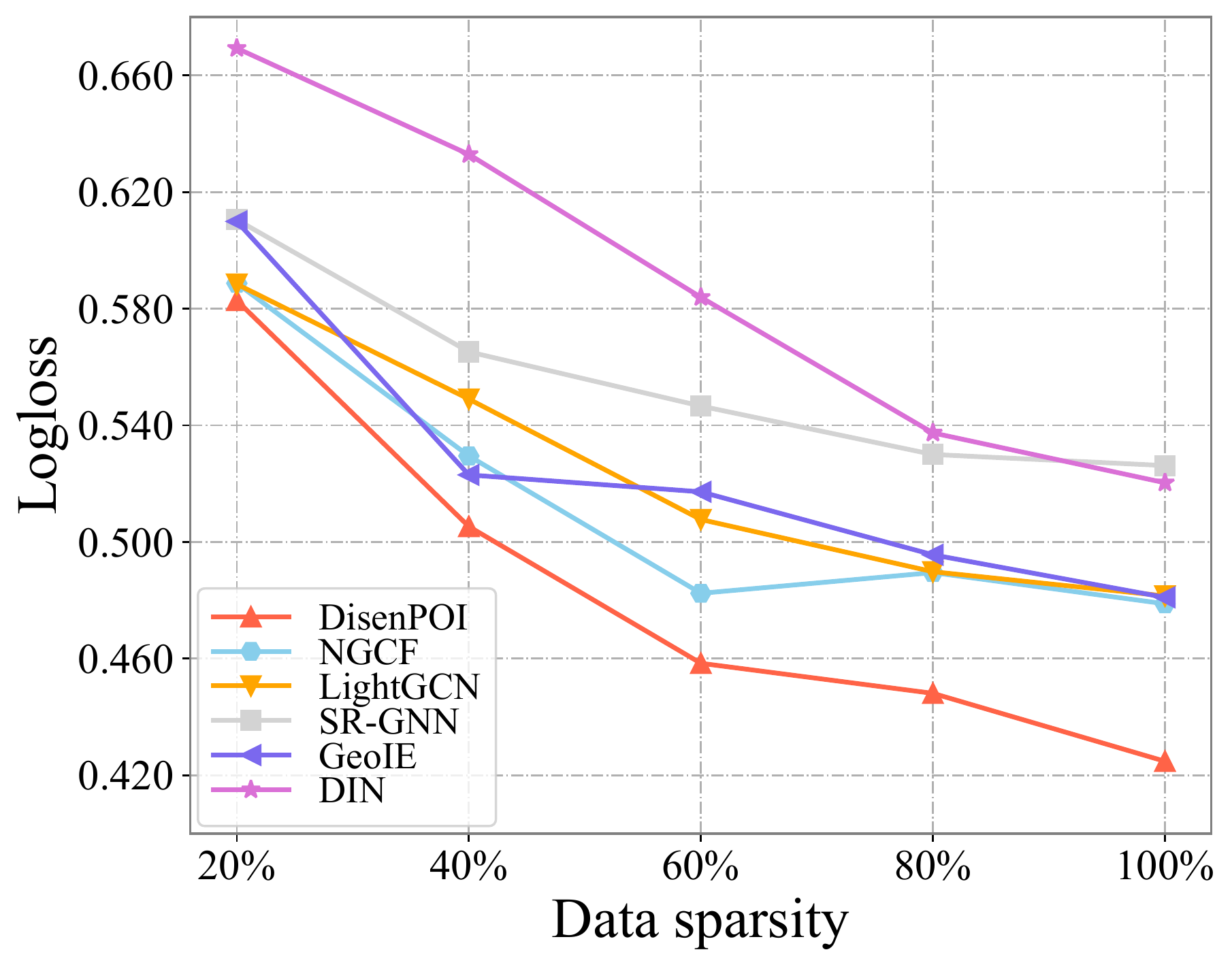}
    \caption{Logloss on \textbf{NYC}}
    \label{fig:coldd}
\end{subfigure}
\caption{Model performance under cold-start settings.} 
\label{fig:cold}
\end{figure}
To investigate whether the idea of disentangling sequential and geographical influence behind visitings can help to alleviate the cold-start issue, we randomly divide each user's visiting sequences into five folds, corresponding to 20\%, 40\%, 60\%, 80\%, and full train set. The experiment results are illustrated in Figure \ref{fig:cold}. We can observe that:
\begin{itemize}[leftmargin=*]
    \item Generally, location-based methods (DisenPOI, GeoIE) can better alleviate the cold-start issue, which indicates that geographical information is essential when dealing with sparse interactions. 
    \item DisenPOI outperforms all baselines under each setting, showing its capability of learning effective representations on sparse data. Specifically, DisenPOI gains more advantages when the training data is rather sparse (under 60\%).
\end{itemize}

\subsection{Analysis of DisenPOI (RQ2)}
\begin{figure}[t]
\centering
\setlength{\abovecaptionskip}{0.1cm}
\setlength{\belowcaptionskip}{-0.3cm}
\begin{subfigure}{0.45\linewidth}
\setlength{\abovecaptionskip}{0.0cm}
\setlength{\belowcaptionskip}{-0.0cm}
    \includegraphics[width=\linewidth]{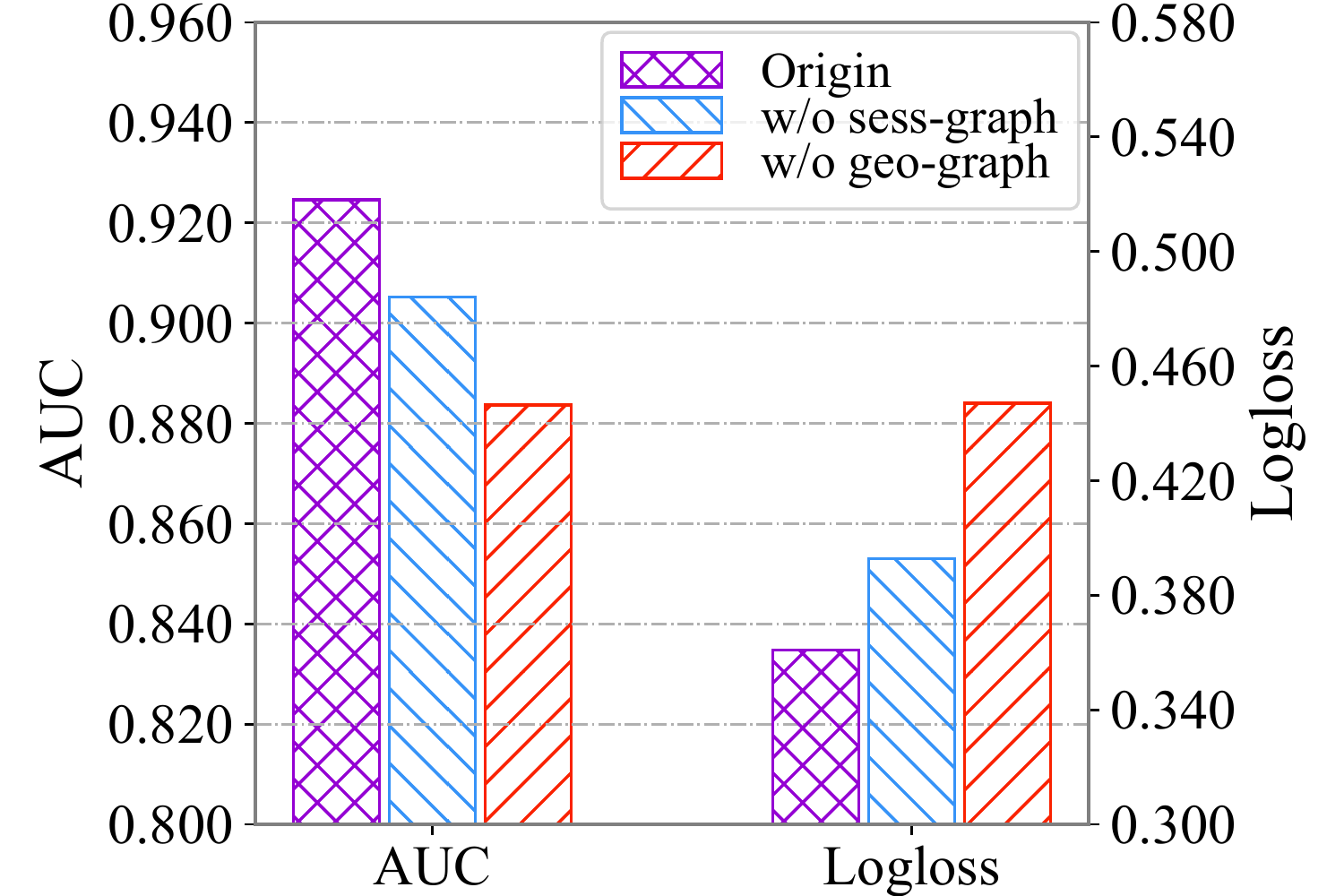}
    \caption{Tokyo}
    \label{fig:1a}
\end{subfigure}
\begin{subfigure}{0.45\linewidth}
\setlength{\abovecaptionskip}{0.0cm}
\setlength{\belowcaptionskip}{-0.0cm}
    \includegraphics[width=\linewidth]{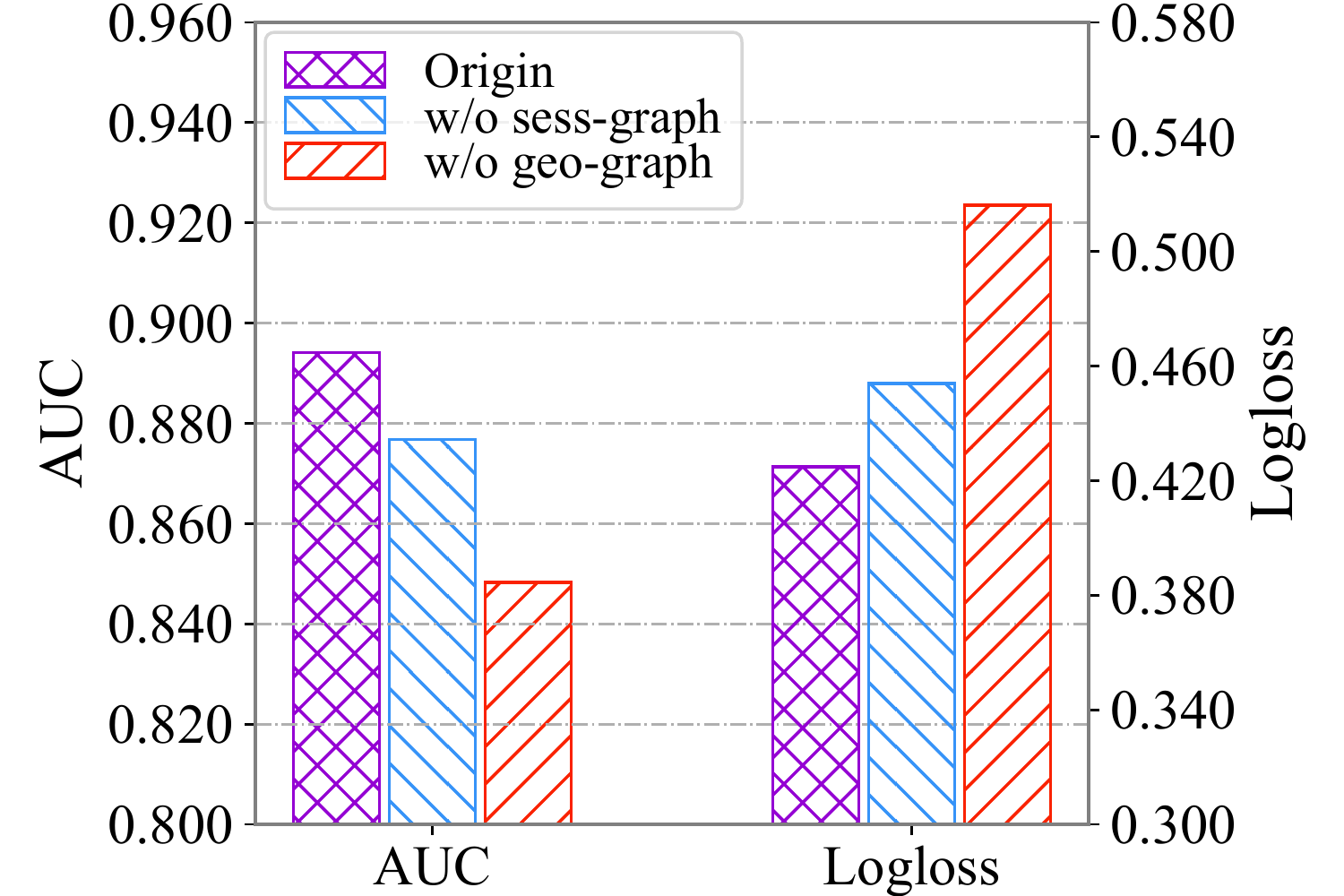}
    \caption{NYC}
    \label{fig:1b}
\end{subfigure}
\begin{subfigure}{0.45\linewidth}
\setlength{\abovecaptionskip}{0.0cm}
\setlength{\belowcaptionskip}{-0.0cm}
    \includegraphics[width=\linewidth]{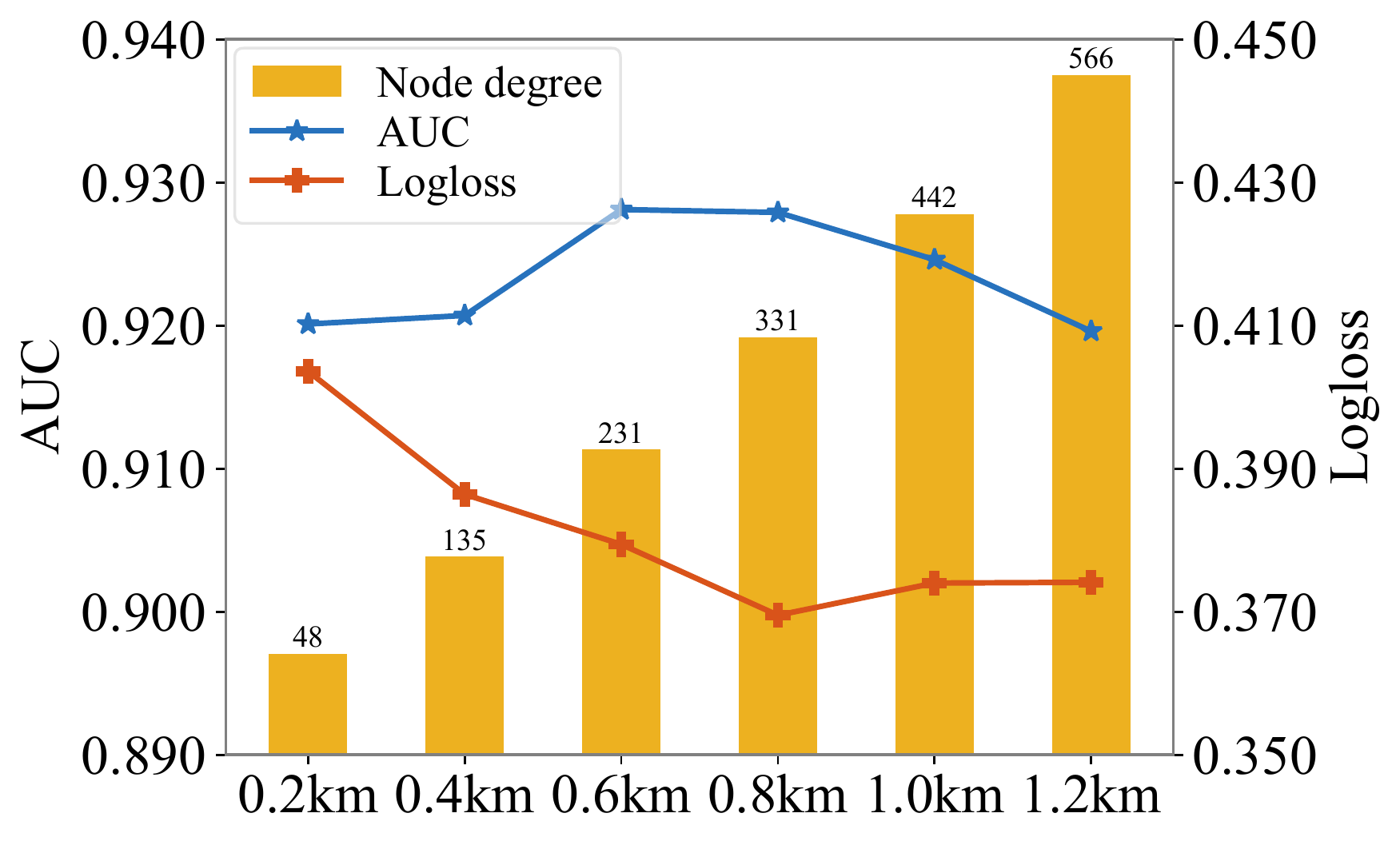}
    \caption{Tokyo}
    \label{fig:1b}
\end{subfigure}
\begin{subfigure}{0.45\linewidth}
\setlength{\abovecaptionskip}{0.0cm}
\setlength{\belowcaptionskip}{-0.0cm}
    \includegraphics[width=\linewidth]{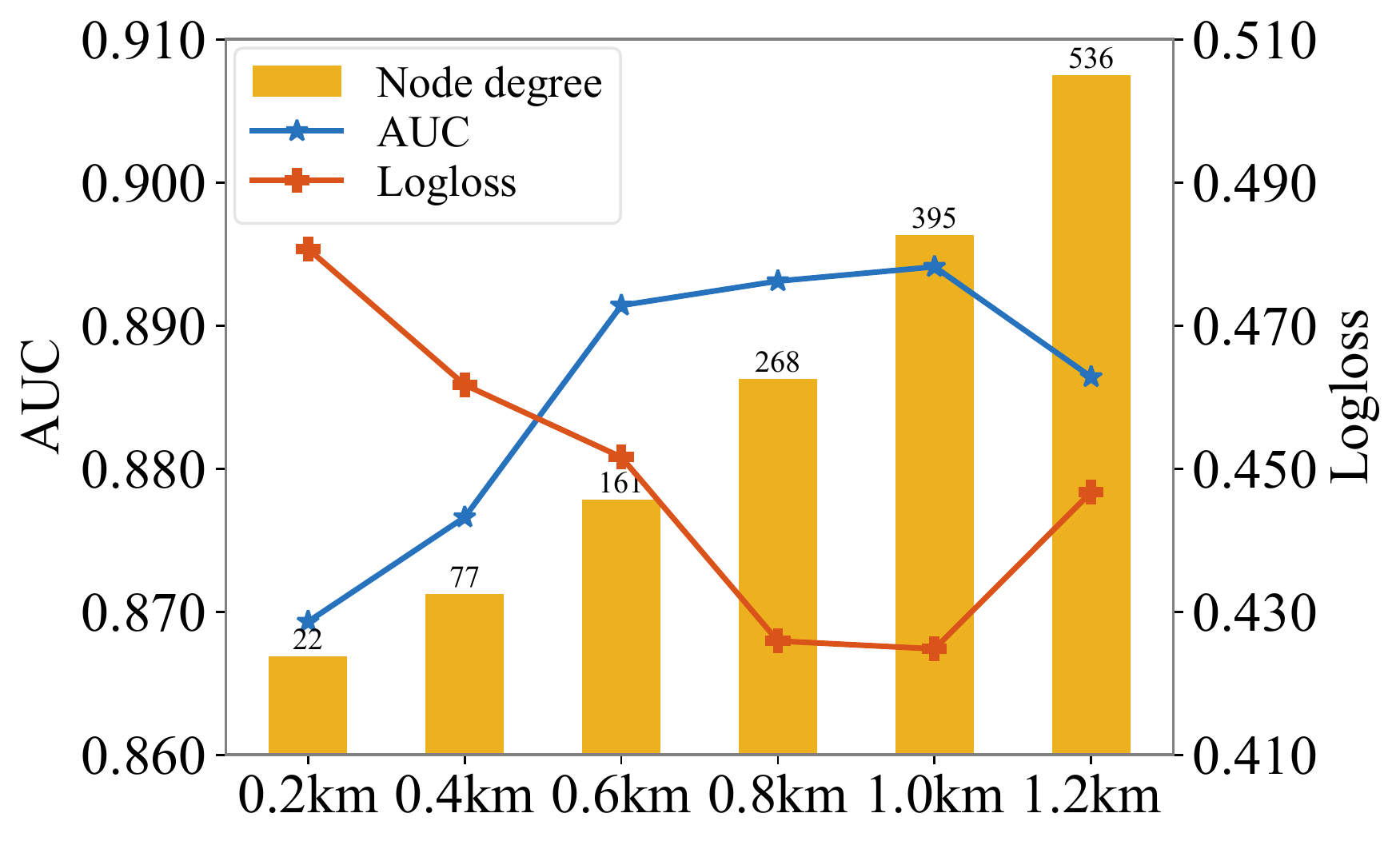}
    \caption{NYC}
    \label{fig:1b}
\end{subfigure}
\caption{Analysis of the dual graph module of DisenPOI.}
\label{fig:1}
\end{figure}
We conduct ablation studies to figure out the functionality and effectiveness of each part in DisenPOI. The following ablation studies are carried out on two \textbf{Foursquare} datasets.
\subsubsection{Effectiveness of Dual Graphs}
To analyze how DisenPOI utilizes the graph data from both geographical and sequential perspectives, we evaluate the model performance under three different settings: the original model, the model without geographical graph (w/o geo-graph), and the model without sequential graph (w/o sess-graph). Since the POI graph $\mathcal{G}_g$ is built based on a specified distance threshold $\Delta d$, we adjust its value between 0.2 km and 1.2 km to analyze the model's sensitivity to $\mathcal{G}_g$. Figure \ref{fig:1} shows:
\begin{itemize}[leftmargin=*]
    \item The outputs of both graph modules play different roles to make recommendations. Whenever removing a type of graph, the model performance declines significantly.
    \item When the distance threshold $\Delta d$ increases, the density of $\mathcal{G}_g$ (average node degree) increases correspondingly, the model performance increase to the optimal and then suffer from a decline. A reasonable $\Delta d$ is 1km, which is close to real world situations.
\end{itemize}
\subsubsection{Difference Between Graph Propagation Methods}
\begin{figure}
\centering
\setlength{\abovecaptionskip}{0.1cm}
\setlength{\belowcaptionskip}{-0.3cm}
\begin{subfigure}{0.51\linewidth}
\setlength{\abovecaptionskip}{0.0cm}
\setlength{\belowcaptionskip}{-0.0cm}
    \includegraphics[width=\linewidth]{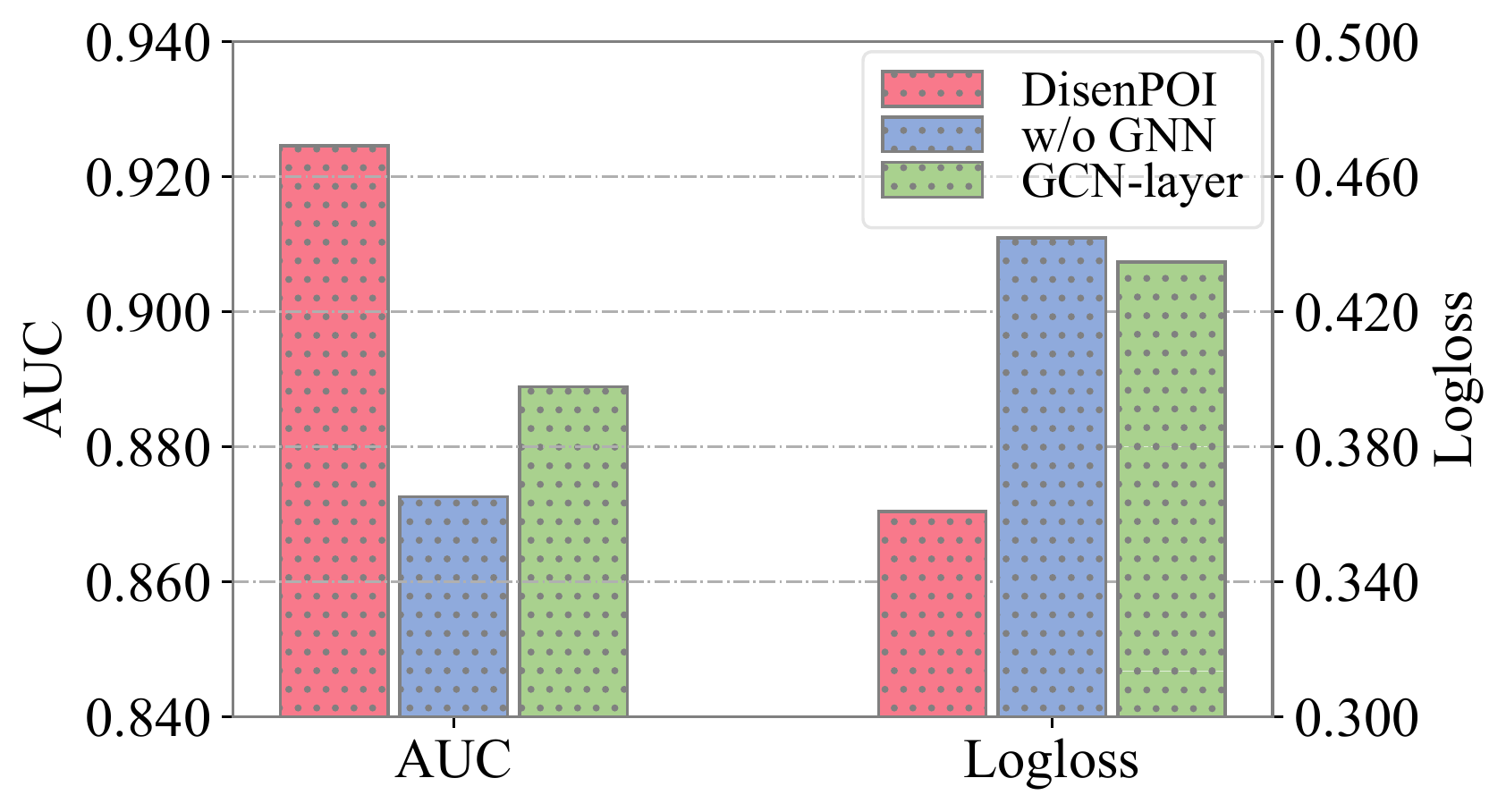}
    \caption{Tokyo}
    \label{fig:2a}
\end{subfigure}
\begin{subfigure}{0.48\linewidth}
\setlength{\abovecaptionskip}{0.0cm}
\setlength{\belowcaptionskip}{-0.0cm}
    \includegraphics[width=\linewidth]{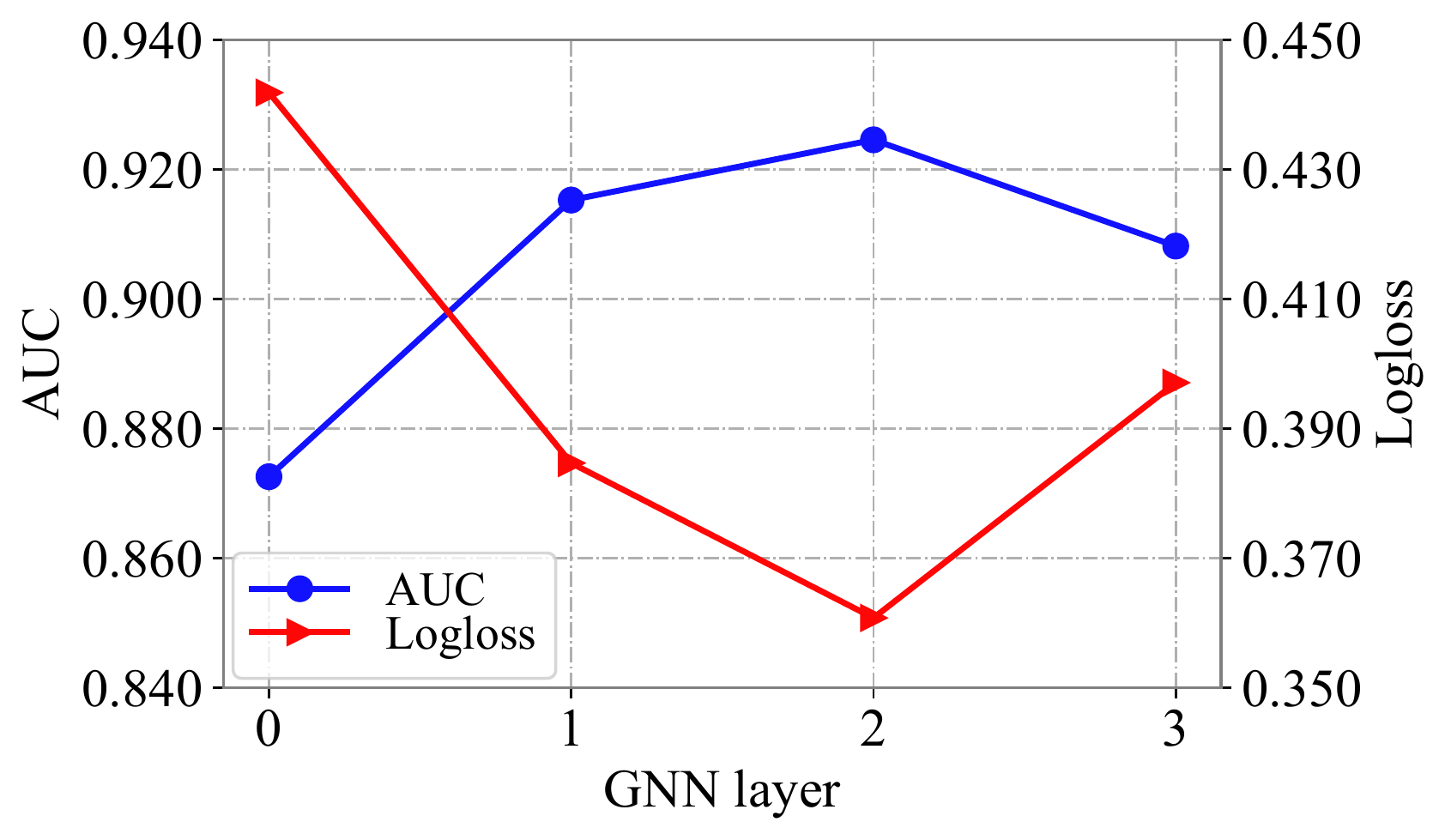}
    \caption{Tokyo}
    \label{fig:2b}
\end{subfigure}

\begin{subfigure}{0.51\linewidth}
\setlength{\abovecaptionskip}{0.0cm}
\setlength{\belowcaptionskip}{-0.0cm}
    \includegraphics[width=\linewidth]{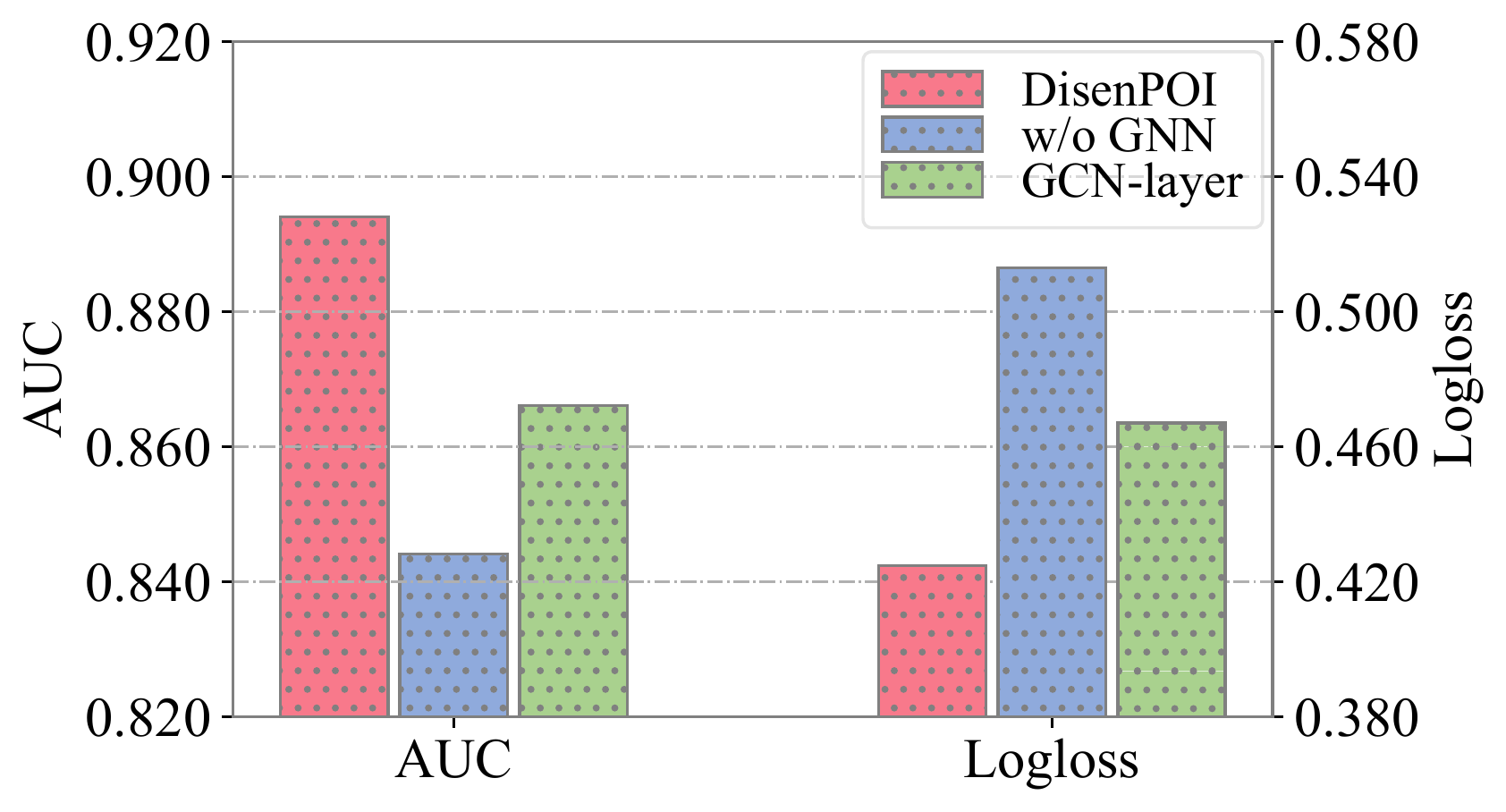}
    \caption{NYC}
    \label{fig:2c}
\end{subfigure}
\begin{subfigure}{0.48\linewidth}
\setlength{\abovecaptionskip}{0.0cm}
\setlength{\belowcaptionskip}{-0.0cm}
    \includegraphics[width=\linewidth]{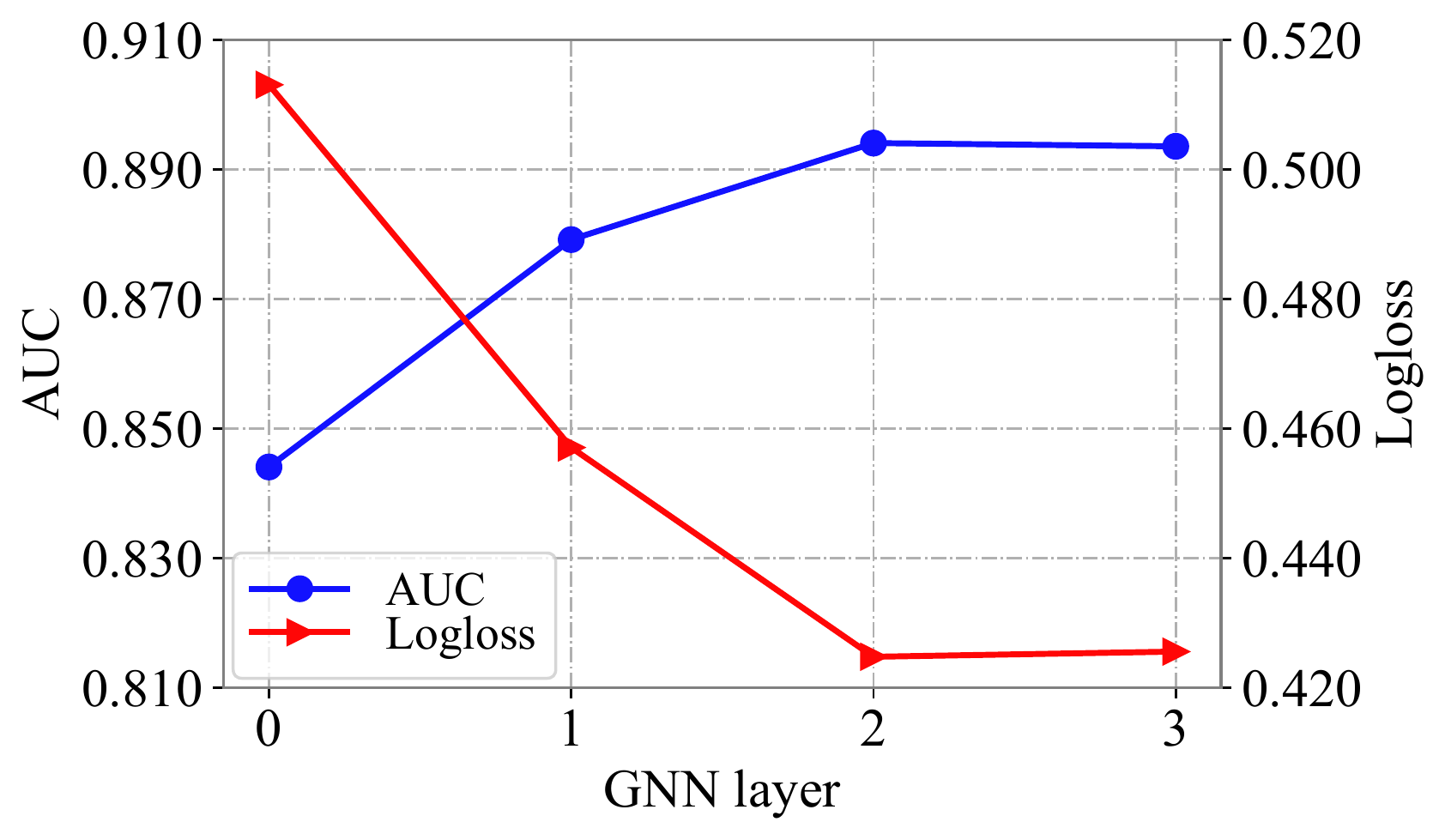}
    \caption{NYC}
    \label{fig:2d}
\end{subfigure}
\caption{Ablation study of GNN layer on \textbf{Foursquare Tokyo}.} 
\label{fig:2}
\end{figure}
As we have proposed two different graph propagation methods for different types of graphs respectively, we investigate the effectiveness of the proposed graph propagation layers. To be specific, we compare model performance when the number of propagation layers is set from 0 (w/o GNN) to 3. We also compare the performance when the GNN layers are replaced by graph convolution networks (i.e. GCN\cite{kipf2016semi}). Based on the results shown in Figure \ref{fig:2}, we have: 
\begin{itemize}[leftmargin=*]
     \item It's important to design specific GNN structures for different graph topologies. Since we adapt two different types of GNNs to better capture the geographical and sequential influence, the model performance would decline once the graph propagation layers are replaced with plain GCNs.
    \item The model achieves its best performance when 2 GNN layers are used, under which setting the model could leverage high-order connectivity of POI nodes on the graph. When there is no GNN layers or the layer amount larger than two, the model would suffer from degenerating or over-smoothing issue of GNNs.
\end{itemize}

\subsubsection{Influence of Disentanglement And Curriculum Learning}
\begin{figure}
\centering
\setlength{\abovecaptionskip}{0.1cm}
\setlength{\belowcaptionskip}{-0.3cm}
\begin{subfigure}{0.49\linewidth}
\setlength{\abovecaptionskip}{0.0cm}
\setlength{\belowcaptionskip}{-0.0cm}
    \includegraphics[width=\linewidth]{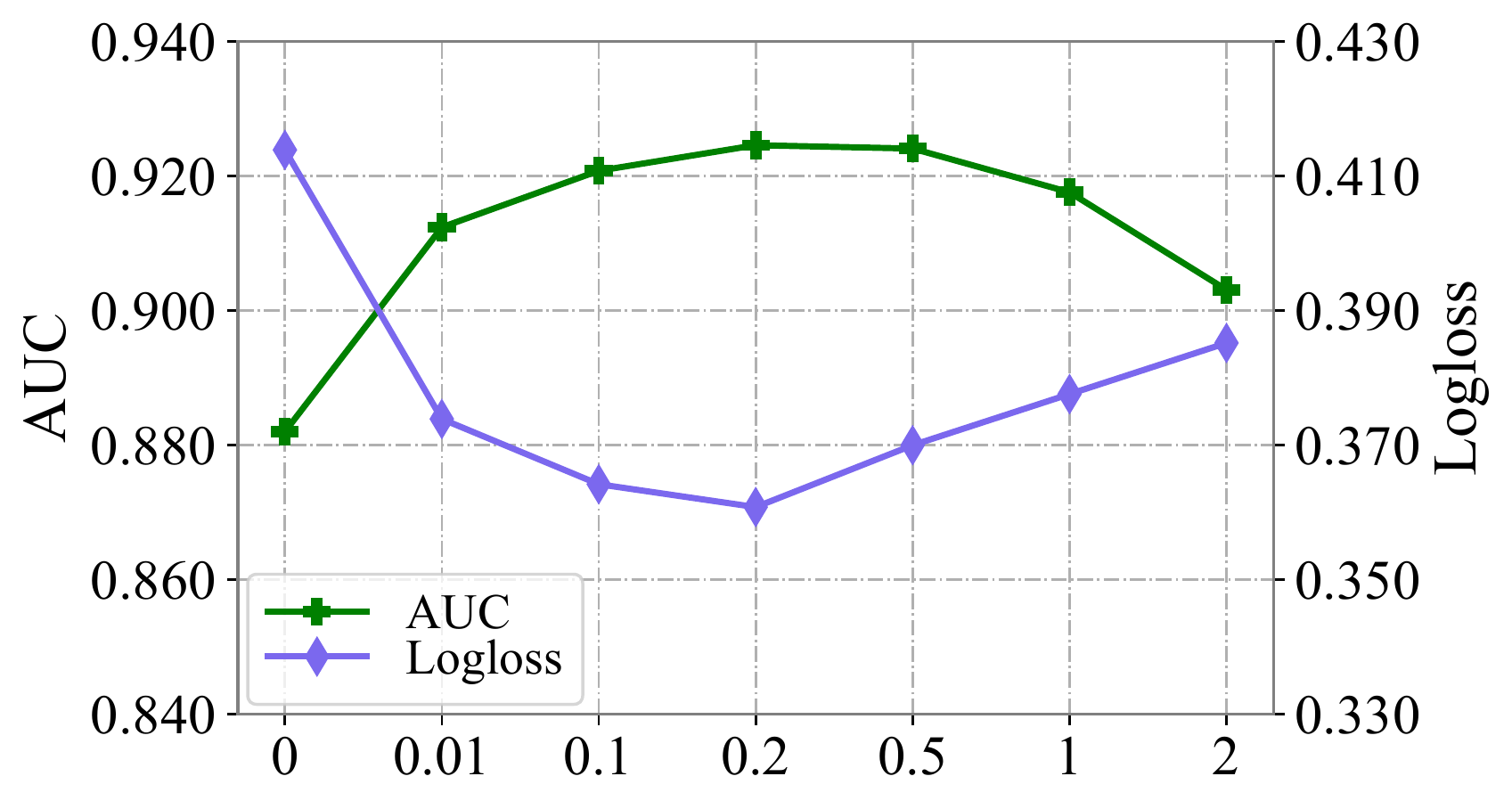}
    \caption{Influence of $\alpha$}
    \label{fig:3a}
\end{subfigure}
\begin{subfigure}{0.49\linewidth}
\setlength{\abovecaptionskip}{0.0cm}
\setlength{\belowcaptionskip}{-0.0cm}
    \includegraphics[width=\linewidth]{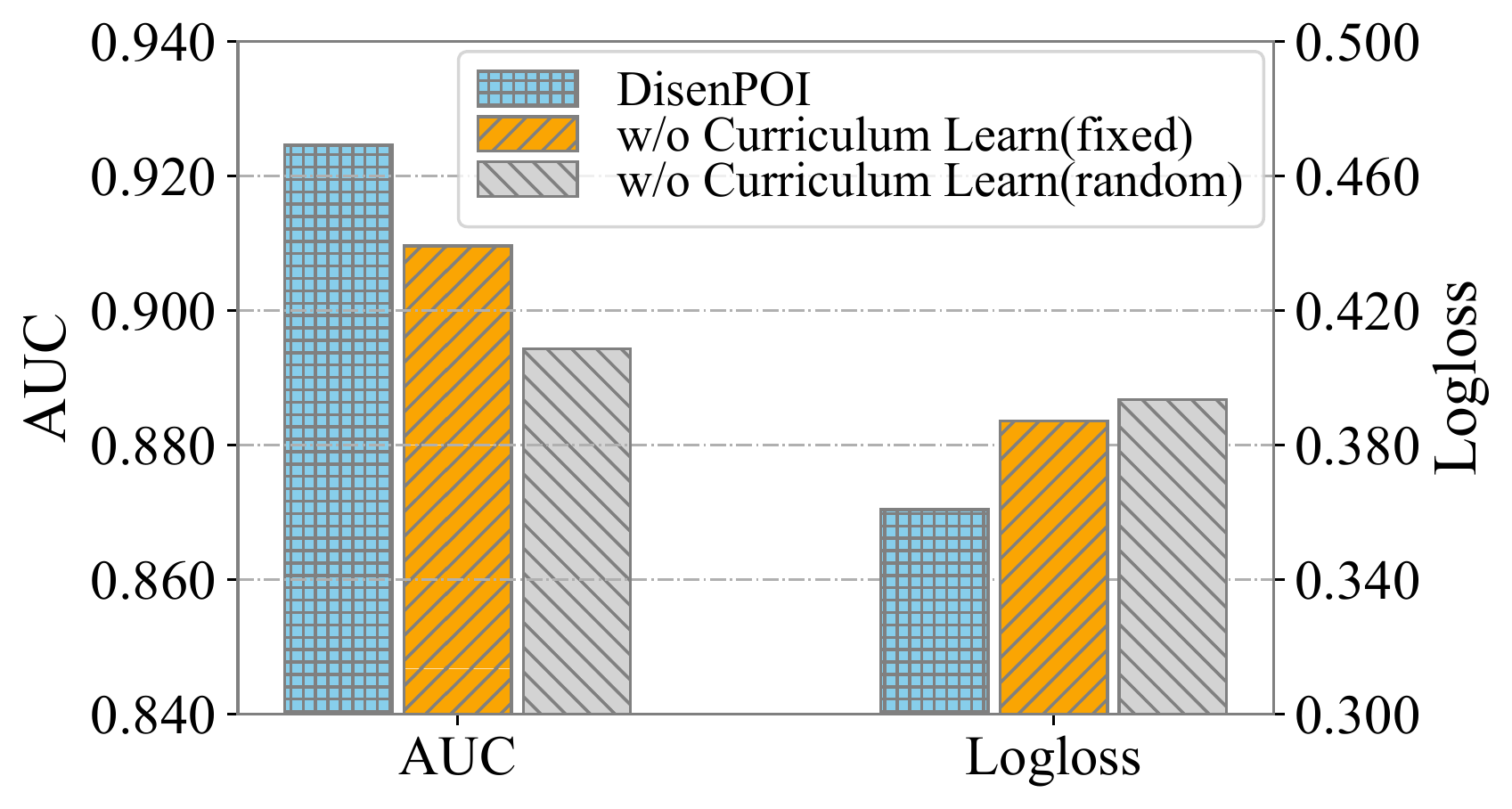}
    \caption{Influce of Curriclum Learning}
    \label{fig:3b}
\end{subfigure}

\caption{Performance under different contrastive settings.} 
\label{fig:3}
\end{figure}
The contrastive loss $\mathcal{L}_{\text{con}}$ in the proposed model serves as a constraint on the model to ensure the similarity between the corresponding graph embeddings and the disentanglement between different types of graph embeddings. To figure out the effectiveness of $\mathcal{L}_{\text{con}}$, we compare the model performance with different contrastive weight $\alpha$ from 0 (w/o contrastive loss) to 2.0. Since the curriculum learning method is applied so that the model converges to a better solution, we compare the model performance under three curriculum settings: our original model, DisenPOI without curriculum learning (with fixed $\alpha$), and DisenPOI with $\alpha$ randomly set under each curriculum step. We conduct the ablation experiments on \textbf{Tokyo} dataset and the results are illustrated in Figure \ref{fig:3}. We observe that:
\begin{itemize}[leftmargin=*]
    \item The contrastive loss improves model performance since the disentangled influence of location and history sequence can be fully utilized to make a prediction. Entangled representations would influence the CTR prediction layer to make a biased or sub-optimal prediction. Generally models with larger $\alpha$ perform better, before suffering with an extreme $\alpha$ (greater than 1.0).
    \item It is necessary to train the model in an easy-to-difficult process to achieve better performance. The models trained with fixed or random $\alpha$ are both likely to converge to a sub-optimal solution instead according to Figure \ref{fig:3b}.
\end{itemize}

\subsection{Visualization and Case Study (RQ3)}
\begin{figure}
\centering
\begin{subfigure}{0.49\linewidth}
\includegraphics[width=\linewidth]{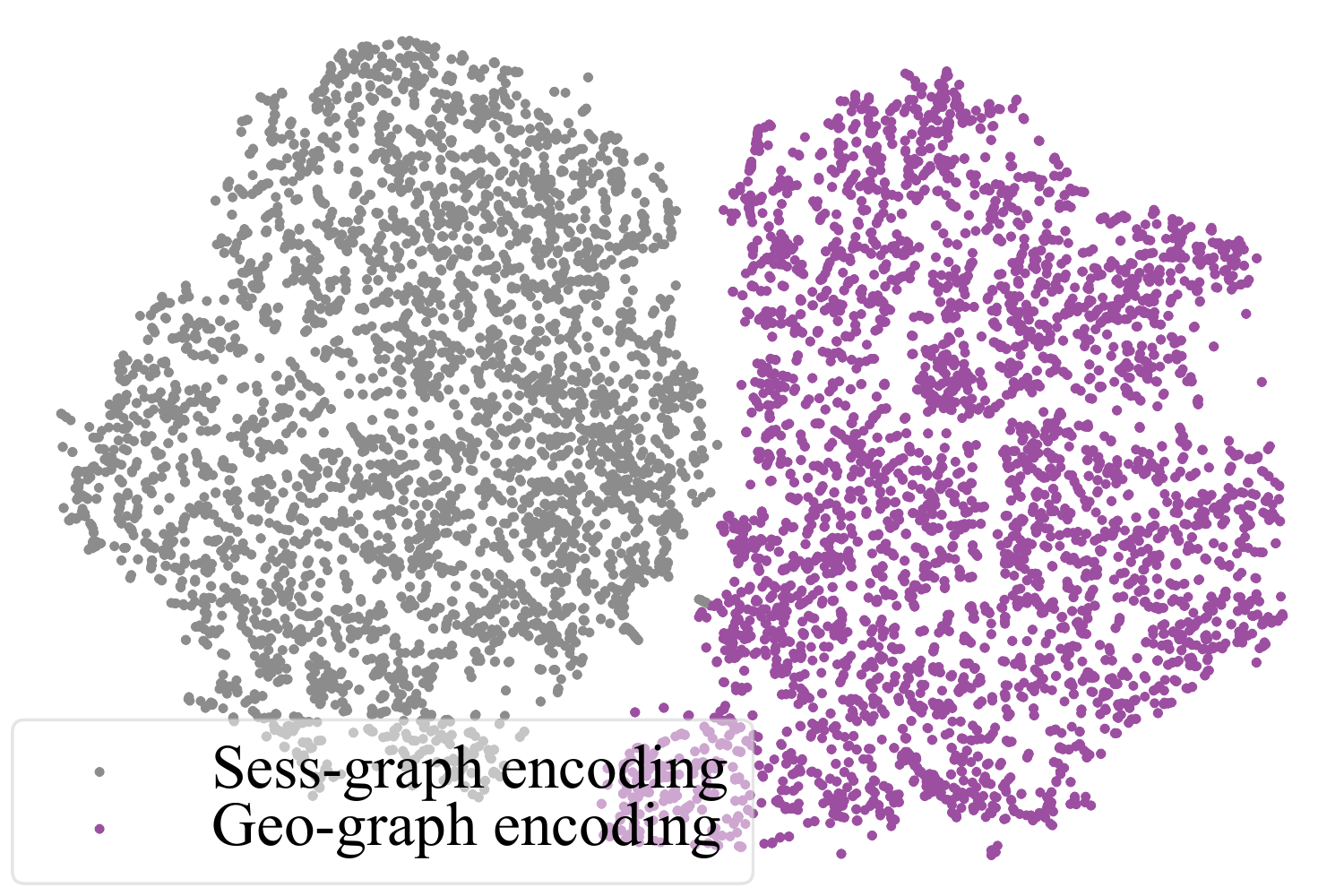}
\caption{Graph Encoding with $\mathcal{L}_{\text{con}}$}
\label{fig:4a}
\end{subfigure}
\begin{subfigure}{0.49\linewidth}
\includegraphics[width=\linewidth]{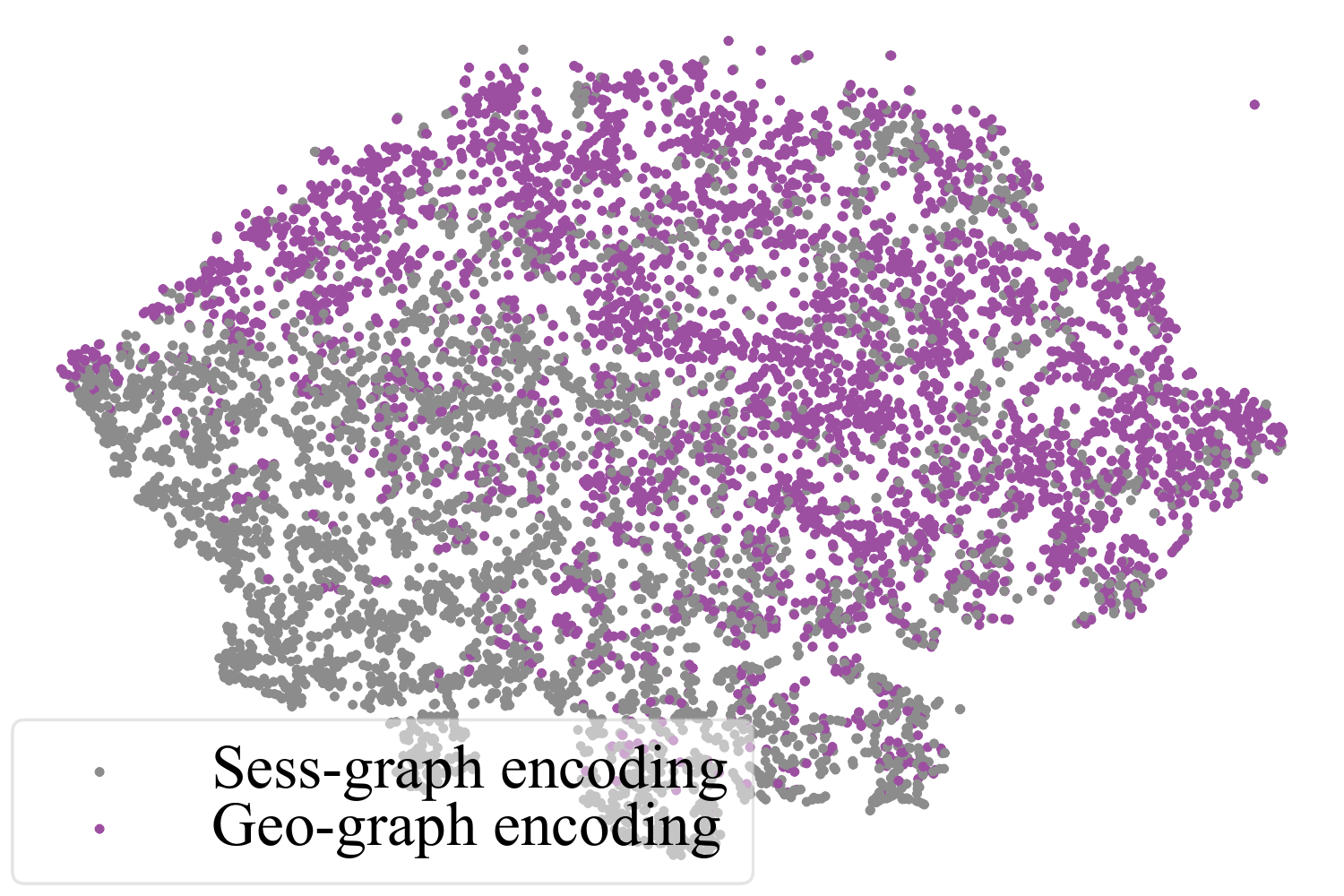}
\caption{Graph Encoding without $\mathcal{L}_{\text{con}}$}
\label{fig:4b}
\end{subfigure}

\begin{subfigure}{0.52\linewidth}
\includegraphics[width=\linewidth]{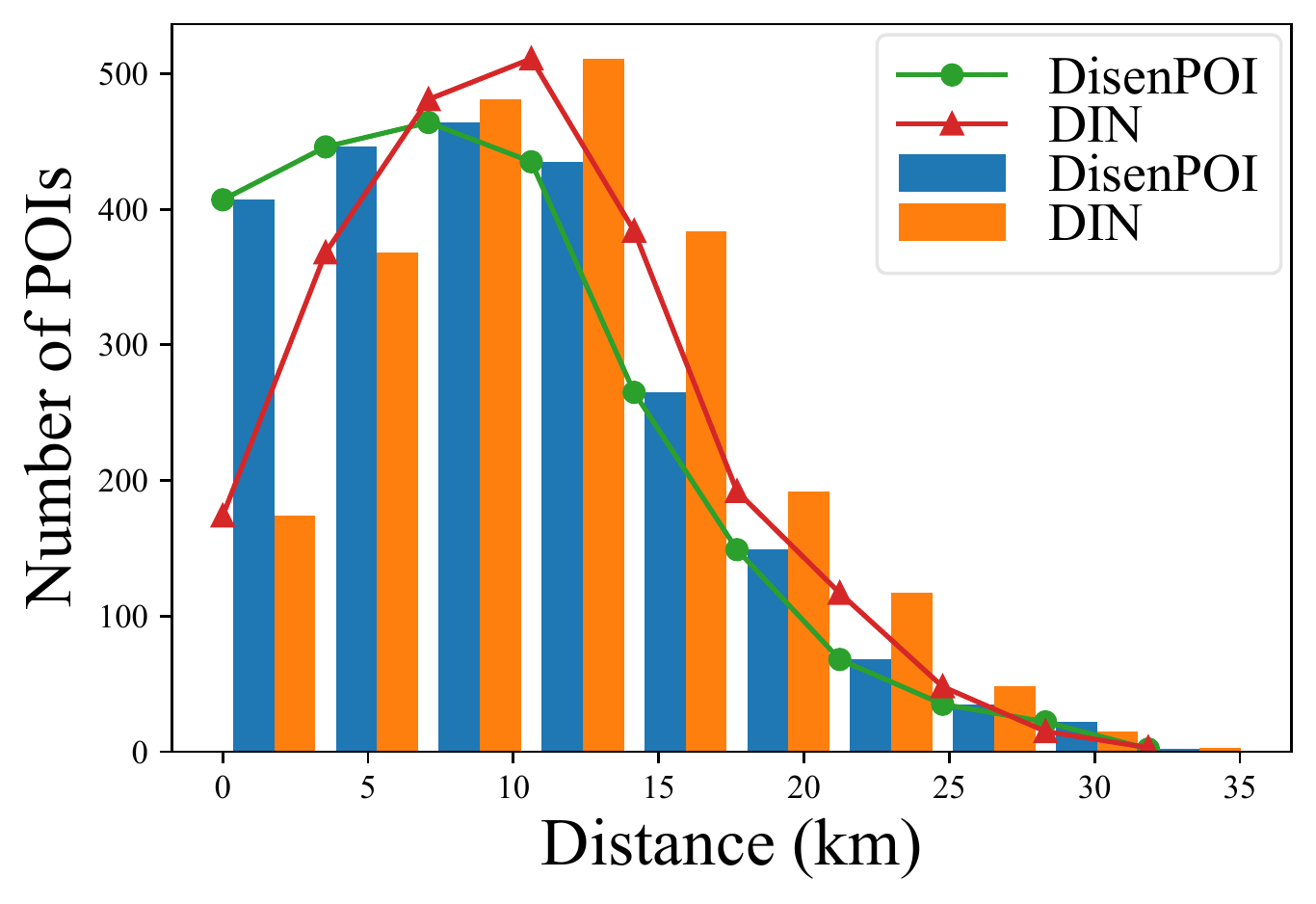}
\caption{Distance distribution of recommended POIs.}
\label{fig:4c}
\end{subfigure}
\begin{subfigure}{0.47\linewidth}
\includegraphics[width=\linewidth]{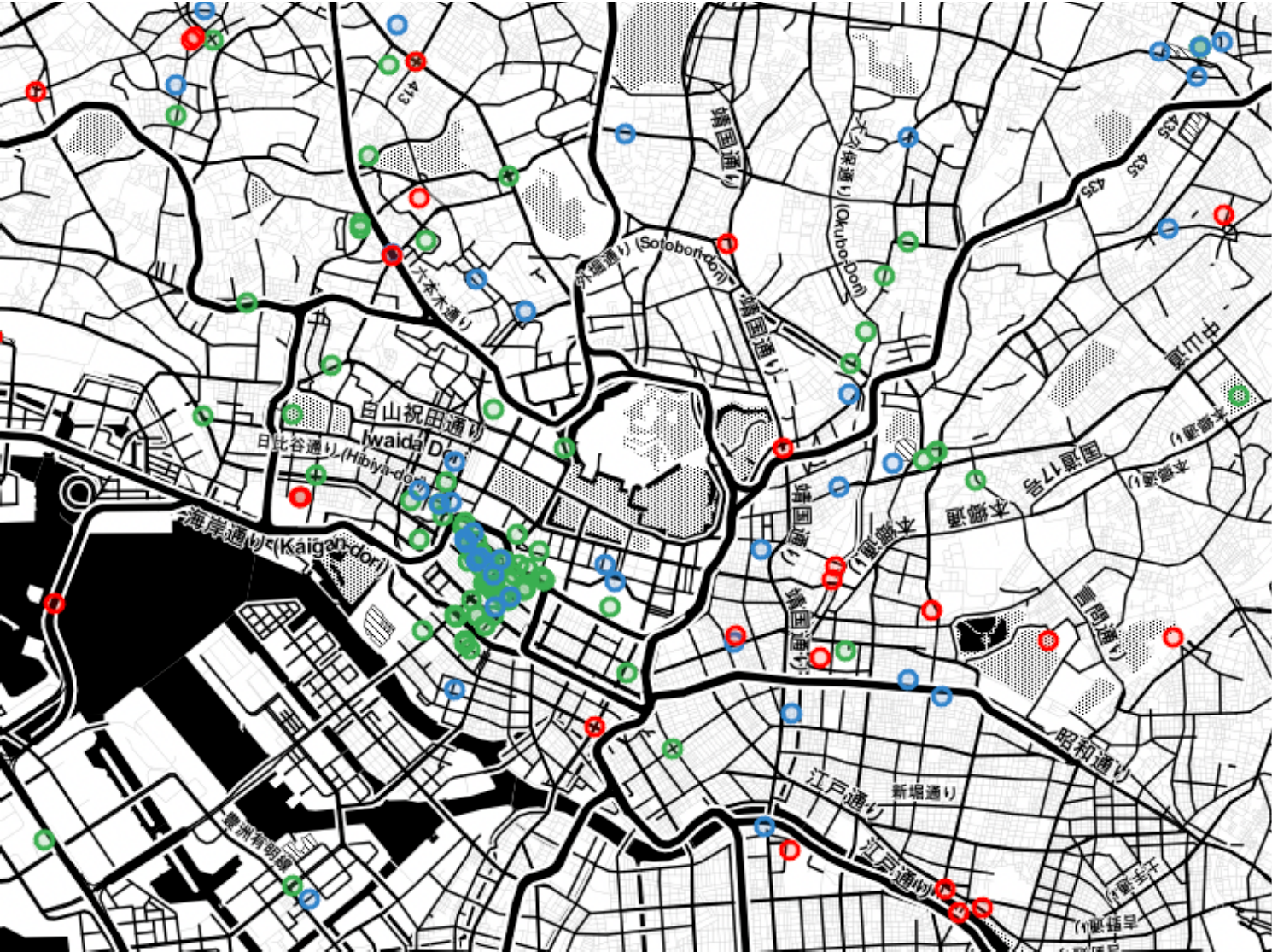}
\caption{User's check-in trajectory.}
\label{fig:4d}
\end{subfigure}

\setlength{\abovecaptionskip}{0.1cm}
\setlength{\belowcaptionskip}{-0.3cm}
\caption{Visualizations of graph encodings and geographical distribution of recommended POIs.}
\label{fig:4}
\end{figure}
\subsubsection{Visualization of Model Output}
To further investigate how the proposed DisenPOI disentangles the hidden representation of POIs, we visualize the representations of user's check-in sequences on the evaluation set. The encodings of visiting sequences are transformed via the t-SNE \cite{van2008visualizing} method. The result shown in Figure \ref{fig:4a} and \ref{fig:4b} illustrates the effect of disentanglement. To be specific, encodings of sequential and geographical graphs form rather clearer cluster structures compared with the model without $\mathcal{L}_{\text{con}}$. The comparison indicates that with the contrastive constraint brought by $\mathcal{L}_{\text{con}}$, DisenPOI successfully clarifies different types of embeddings and encodes two graphs into disentangled latent spaces.

To demonstrate the effectiveness of the proposed location-based idea, we illustrate in Figure \ref{fig:4c} by comparing the distance between target POIs in the evaluation set and the predictions made by DisenPOI and DIN. It shows that DisenPOI does a better job in recommending POIs that are close to the target than traditional methods.

\subsubsection{Case Study}
\begin{table}
\centering
\setlength{\abovecaptionskip}{-0cm}

\caption{Difference of top recommended POIs.}
\label{tab:3}
\begin{tabular}{cc|cc} 
\toprule 
\multicolumn{2}{c}{\textbf{DisenPOI}} & \multicolumn{2}{c}{\textbf{DIN}} \\
\textbf{Distance(km)} & \textbf{Category} & \textbf{Distance(km)} & \textbf{Category} \\
\midrule 
0.310 & Road & 4.727 & Train Station\\
1.331 & Café & 3.908 & Subway \\
0.780 & Smoke Shop & 0.981 & Art Museum \\
2.971 & Subway & 5.800 & Bridge\\
0.462 & Gift Shop & 2.544 & Train Station\\
\midrule
\textbf{Average} & \textbf{ACC} & \textbf{Average} & \textbf{ACC} \\
2.41 & 29.6\% & 5.97 & 28.3\%\\
\bottomrule 
\end{tabular}
\end{table}
We randomly choose a user (488) who has a visiting sequence with 148 POIs for detailed investigation. We compare the top-100 recommended POIs of DisenPOI and DIN for the selected user (488) in Figure \ref{fig:4d}. POIs visited by the user are marked with green circles, while POIs recommended by DisenPOI and DIN are marked with blue and red circles respectively. It can be observed that the POIs recommended by DisenPOI are generally closer to the selected user's visiting trajectory than POIs recommended by DIN, which shows the effectiveness of modeling location-based influence with a geographical POI graph.

To make an intuitive comparison, we compare the top scored POIs recommended by DisenPOI and DIN in Table \ref{tab:3}. We first list the distance and categories of top-5 recommended POIs, then the average distance and accuracy on the category of top-50 recommended POIs. It can be observed that DIN pays more attention to other features of the target POI, such as its category (e.g. train station) and functionality (e.g. transportation), while DisenPOI considers both geographical influence and POI's functionality when making recommendations. 
The results shows that the ignorance of geographical effects will cause models to recommend POIs that share similarities with user interests but at a rather far distance, which is less likely to be visited next. By contrast, DisenPOI would prefer a similar POI within a short distance from user's current location, which helps it to make more reasonable recommendations.


%% file: conclusion.tex
\section{Conclusion}
In this paper, we propose DisenPOI, a dual-graph model that disentangles sequential and geographical influence for Point-of-Interest recommendations.
We construct a pair of a sequence-based graph $\mathcal{G}_{s,u}$ and a location-based graph $\mathcal{G}_{g,u}$ to model the disentangled sequential and geographical factor behind a visit. We use customized graph propagation schemes for accurate delineation of different semantics on both graphs and leverage contrastive loss to achieve disentanglement. Experiments on three datasets demonstrate the effectiveness and expressiveness of DisenPOI to disentangle sequential and geographical influence on location-based CTR prediction.